\documentclass{revtex4}
\usepackage{amsfonts}
\usepackage{amsmath}
\usepackage{amssymb}
\usepackage{version}
\usepackage{graphicx}

\includeversion{New_connection}
\excludeversion{Old_connection}

\begin{document}

\title[RCSJ in the age of MQT]{A survey of classical and quantum
interpretations of experiments on Josephson junctions at very low
temperatures}
\author{James A. Blackburn}
\affiliation{Physics \& Computer Science, Wilfrid Laurier University, Waterloo, Ontario,
Canada}
\author{Matteo Cirillo*}
\affiliation{Dipartimento di Fisica and MINAS-Lab, Universit\`{a} di Roma
\textquotedblleft Tor Vergata\textquotedblright\ I-00133 Roma, Italy}
\author{Niels Gr\o nbech-Jensen}
\affiliation{Department of Mathematics, University of California, Davis, CA 95616\\
Department of Mechanical and Aerospace Engineering, University of
California, Davis, CA 95616}

\begin{abstract}
For decades following its introduction in 1968, the resistively and
capacitively shunted junction (RCSJ) model, sometimes referred to as the
Stewart-McCumber model, was successfully applied to study the dynamics of
Josephson junctions embedded in a variety of superconducting circuits. In
1980 a theoretical conjecture by A.J. Leggett suggested a possible new and
quite different behavior for Josephson junctions at very low temperatures. A
number of experiments seemed to confirm this prediction and soon it was
taken as a given that junctions at tens of millikelvins should be regarded
as macroscopic quantum entities. As such, they would possess discrete levels
in their effective potential wells, and would escape from those wells (with
the appearance of a finite junction voltage) via a macroscopic quantum
tunneling process. A zeal to pursue this new physics led to a virtual
abandonment of the RCSJ model in this low temperature regime. In this paper
we consider a selection of essentially prototypical experiments that were
carried out with the intention of confirming aspects of anticipated
macroscopic quantum behavior in Josephson junctions. We address two
questions: (1) How successful is the non-quantum theory (RCSJ model) in
replicating those experiments? (2) How strong is the evidence that data from
these same experiments does indeed reflect macroscopic quantum behavior?

(*) Also \textit{CNR-SPIN} Institute, Italy
\end{abstract}

\maketitle

\newpage

\tableofcontents

\newpage 

\section{\protect\large Introduction}

The history of the Josephson junction began with Brian Josephson's paper in
1962 \cite{Josephson}. The 50th anniversary of this discovery was marked at
the 2012 Applied Superconductivity Conference in Portland OR.

\begin{figure}[t]
\begin{center}
\scalebox{0.3}{\centering \includegraphics{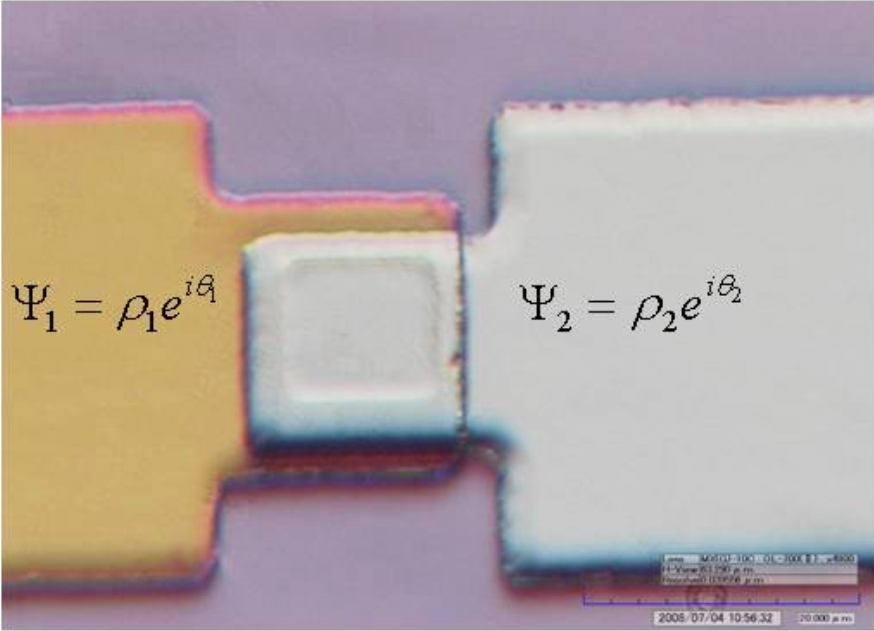}}
\end{center}
\vspace{-0.5 cm}
\caption{Photomicrograph of a thin film Josephson junction. \ The small
square area at the center ($10\protect\mu m\times 10\protect\mu m$) is the
active junction. \ The superconducting order parameters $\Psi $ for each
film are indicated.}
\label{jj}
\end{figure}

\begin{figure}[t]
\begin{center}
\scalebox{0.3}{\centering \includegraphics{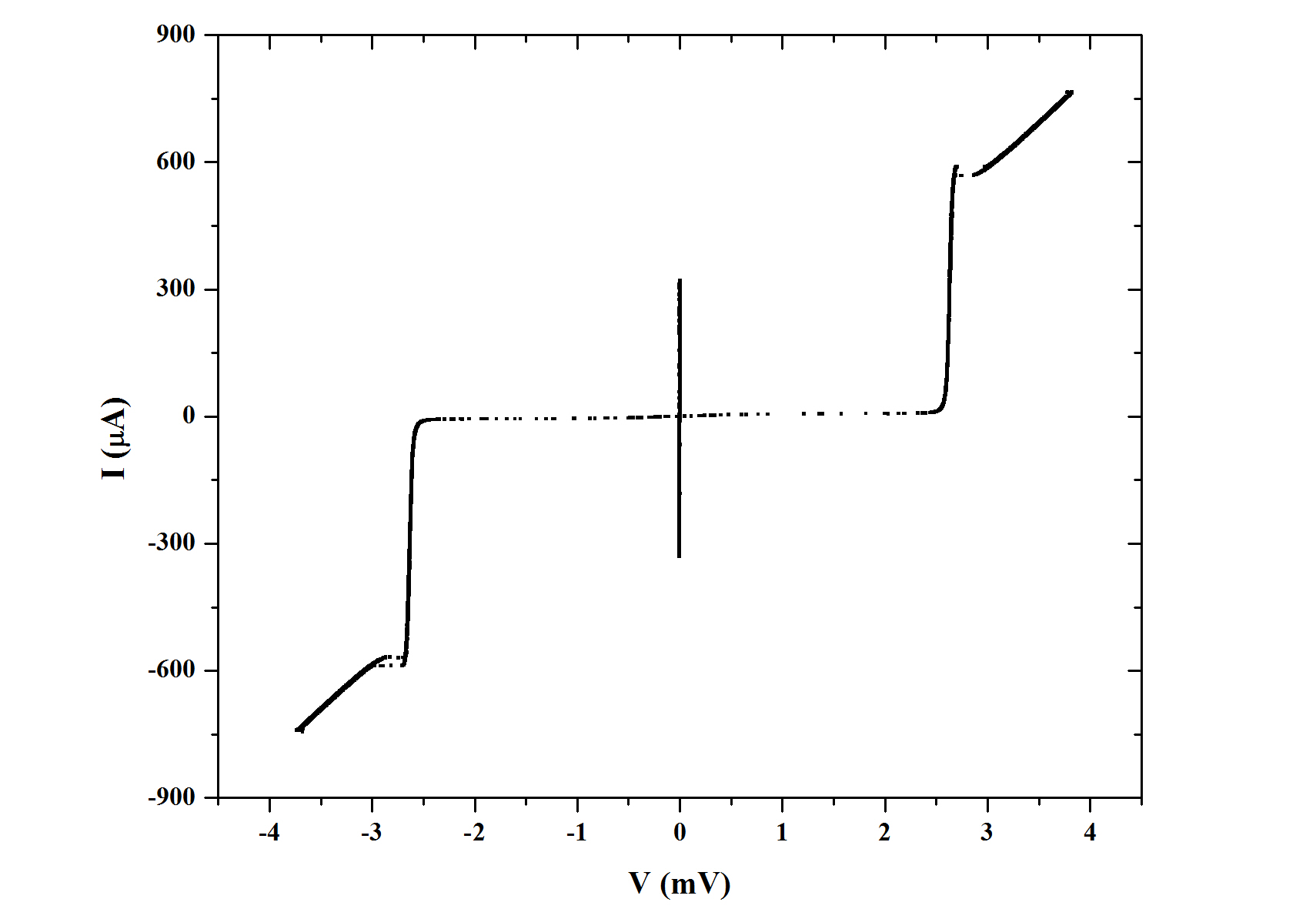}}
\end{center}
\vspace{-0.5 cm}
\caption{Typical current-voltage
characteristic for an underdamped Josephson junction.}
\label{IV}
\end{figure}

At its simplest, a Josephson junction consists of two superconductors
separated by a very thin barrier, typically an oxide of a superconducting
material. Figure \ref{jj} shows a typical thin film junction. The Nb
electrodes each have a small finger, and these fingers overlap to form a
Josephson junction from the sandwich of Nb top and bottom, and the oxide in
between. In the most successful fabrication technology \cite{Gurwitch83} the
oxide is $Al_{2}O_{3}$ thermally grown over a thin aluminum layer wetting
the Nb bottom layer. The two bulk superconducting electrodes can be viewed
phenomenologically as possessing separate macroscopic wavefunctions $\Psi
_{1}$ and $\Psi _{2}$, as indicated in the figure, whose phase difference $%
\theta _{1}-\theta _{2}$ is $\varphi $. Josephson predicted that a
supercurrent could flow through the junction without any associated voltage
- the expression for this supercurrent being
\begin{equation}
I=I_{C}\sin \varphi  \label{Josephson}
\end{equation}%
If a voltage does exist across the junction, then the phase is governed by 
\begin{equation}
d\varphi /dt=2eV/\hbar  \label{phase}
\end{equation}%
For small junctions, those two equations comprise a complete specification
of the junction dynamics. A typical current-voltage characteristic of a $%
Nb-Al_{2}O_{3}-Nb$ Josephson tunnel junction having a zero-voltage
supercurrent of few hundreds microamp\`{e}res is shown in Fig. \ref{IV}. 

\begin{figure}[t]
\begin{center}
\scalebox{1.0}{\centering \includegraphics{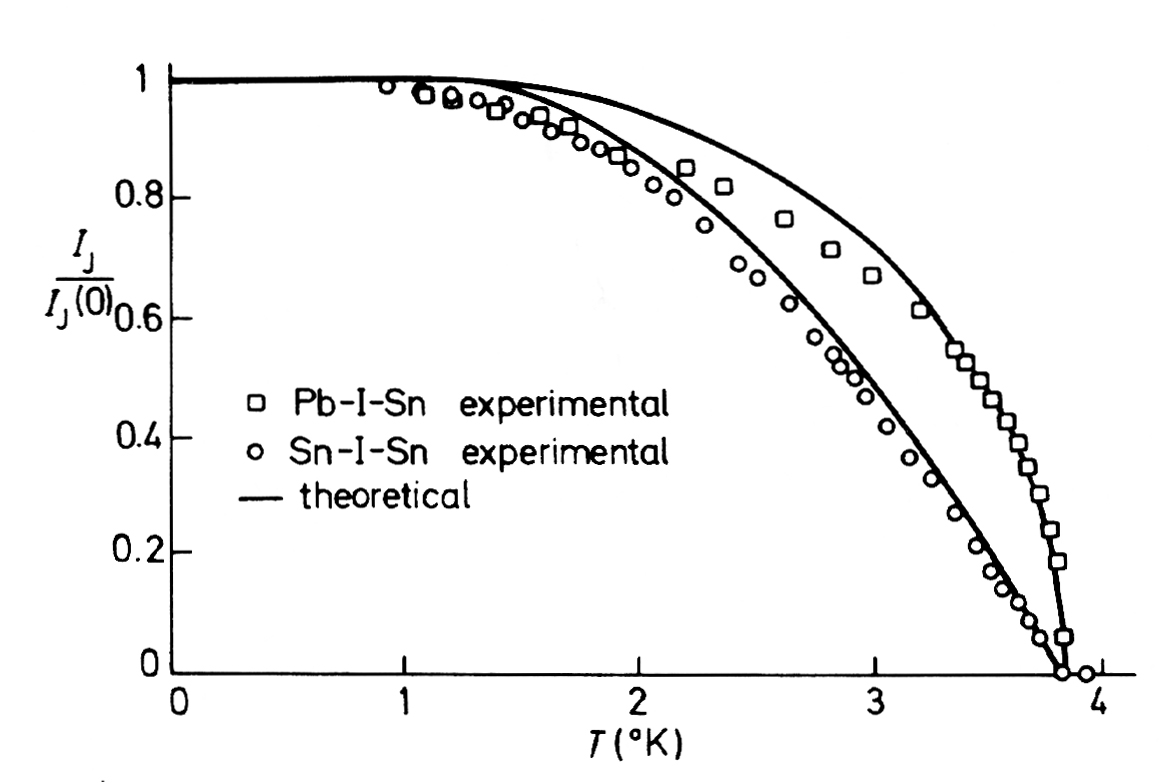}}
\end{center}
\vspace{-0.5 cm}
\caption{Variation of maximum supercurrent in a Josephson
junction as a function of temperature.}
\label{Solymar}
\end{figure}

A surprising fact implied by the above equations is that the existence of
the Josephson effect requires only a pair of superconducting electrodes -
the choice of specific materials plays no role other than to affect the
critical current $I_{C}$. As illustrated in Fig.\ref{Solymar}, this critical
current rises as the junction temperature drops and saturates at low
temperatures \cite{Solymar}. The data in this plot are taken from \cite%
{Fiske} and the theoretical curves are from \cite{Ambegaokar}. 
Therefore, well below the transition temperature of the films
there would be no additional temperature dependence in a Josephson device.
Relevant and fundamental observations on the nature of the Josephson phase
difference and associated energy/potential were reported by P. W. Anderson 
\cite{PWA64}

In 1980 Leggett \cite{Leggett} postulated that , at sufficiently low
temperatures, the phase of a Josephson junction would take on the attributes
of a macroscopic quantum coordinate. Since then, over the following decades,
experimental and theoretical work has been reported demonstrating evidence
that a quantum interpretation of experiments performed at low temperatures
on Josephson junctions is possible.\ Later we will duly reference and
analyze several of these experiments which are complex and sophisticated and
were performed bearing in mind the idea that at very low temperatures a
macroscopic quantum description was necessary.

It is our opinion, however, that \textbf{decisive} confirmation of the
macroscopic quantum hypothesis requires an experiment (or many experiments)
whose results \textbf{cannot be explained} in other terms. In particular, it
is known that there exists (and already existed in the early 80's indeed) a
whole field of research on nonlinear circuits and applied superconductivity
in which the Josephson phase is treated as a classical continuum variable in
the sense specified by P. W. Anderson \cite{PWA64}. We will herein refer to
this approach as the "classical" approach. Thus the necessity of new quantum
interpretation of experiments would be mandatory in case these experiments
cannot be explained by the classical approach. In this Review we consider a
number of key experiments that were initially claimed to unequivocally
verify the quantum hypothesis. We use data extracted from those publications
to very carefully reconsider the matter using the classical Josephson
junction model applied to the specific conditions of each experiment.

In the next section we review shortly the features of the classical model of
Josephson junctions and circuits and successively we turn to a set of
essentially defining experiments and examine in detail the evidence for and
against both the classical and quantum interpretations.

\section{\protect\large The Classical Approach}

\subsection{RCSJ Junction Model}

\begin{figure}[t]
\begin{center}
\scalebox{0.5}{\centering \includegraphics{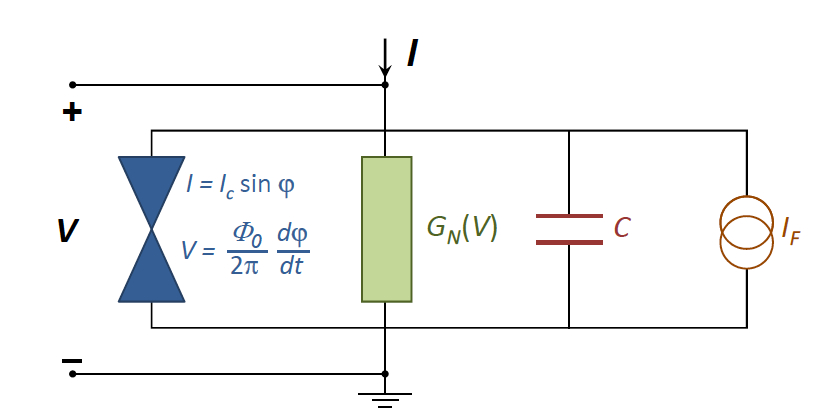}}
\end{center}
\vspace{-0.5 cm}
\caption{Resistively and capacitively
shunted model of a Josephson junction (RCSJ).}
\label{RCSJ}
\end{figure}

In the 1968 a very simple equivalent circuit, depicted in Fig.\ref{RCSJ},
for a Josephson junction was proposed by W.J. Johnson (pp122-124, \cite%
{Barone}), D.E. McCumber, and W.C. Stewart, \cite{Stewart, McCumber,VanDuzer}.
This circuit \cite{Gross}
captures the essential ingredients of a real Josephson device. It is
comprised of three parallel elements: a shunt resistor R, a shunt capacitor
C, and a pure Josephson element usually labeled here with a bowtie graphic
-- the RCSJ model. The RCSJ equivalent circuit for a Josephson junction has
proven to be exceptionally successful in modeling the dynamics of the
Josephson systems.

The current through a Josephson element is given by $I$ and the voltage
across the parallel combination is governed by $d\varphi /dt=2eV/\hbar $.
With a total applied bias current $I$, the phase dynamics are governed by

\begin{equation}
\frac{\hbar C}{2e}\frac{d^{2}\varphi}{dt^{2}}+\frac{\hbar}{2eR}\frac{%
d\varphi }{dt}+I_{C}\sin\varphi=I  \label{Eq1}
\end{equation}

If time is normalized to $1/\omega _{J0}$ where $\omega _{J0}=\sqrt{%
2eI_{C}/\hbar C}$ is the zero-bias Josephson plasma frequency, then

\begin{equation}
\ddot{\varphi}+\alpha \dot{\varphi}+\sin \varphi =\eta  \label{Eq2}
\end{equation}%
where $\eta =I/I_{C}$ is a normalized bias current and $\alpha =\frac{\hbar
\omega _{J0}}{2eI_{C}R}$ is a normalized loss coefficient.

A Josephson junction with phase $\varphi $ has stored energy is $\frac{\hbar
I_{C}}{2e}\left( 1-\cos \varphi \right) $. The pre-factor in this expression
is the Josephson energy:

\begin{equation}
E_{J}=\hbar I_{C}/2e  \label{Eq3}
\end{equation}

\subsection{Washboard Potential}

The total potential energy of a junction, when an additional bias current is
supplied, is \cite{PWA64}

\begin{equation}
U=E_{J}\left\{ \left( 1-\cos\varphi\right) -\eta\varphi\right\}  \label{Eq4}
\end{equation}

\begin{figure}[t]
\begin{center}
\scalebox{0.25}{\centering \includegraphics{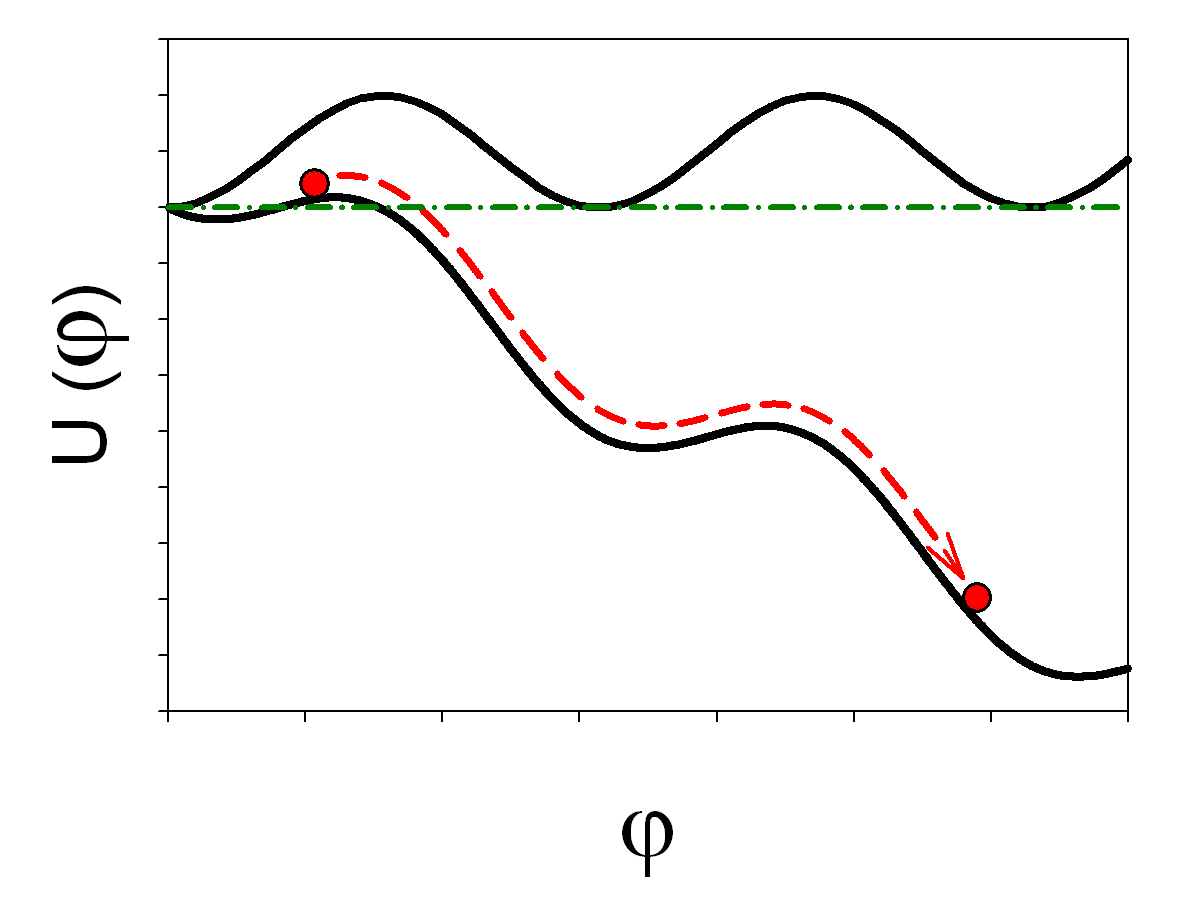}}
\end{center}
\vspace{-0.5 cm}
\caption{Washboard potential with zero
bias (upper) and non-zero bias (lower).}
\label{Fig1}
\end{figure}

In this form it is apparent that the phase dynamics can be viewed in terms
of a fictitious `particle' whose coordinate is $\varphi$, moving in a
washboard potential, as indicated in Fig.\ref{Fig1}.

In this analog, at zero bias the potential is a horizontal washboard and a
`particle' would sit at the bottom of the well at $\varphi =0$. \ Small
oscillations around the minimum of that well occur at the plasma frequency $%
f_{J0}=\omega _{J0}/2\pi $. At non-zero values of the bias: (1) the
washboard tilts (2) the minimum of the well occupied by the particle shifts
to $\varphi >0$ (3) the wells in the washboard potential become
progressively shallower, with correspondingly smaller plasma frequencies, $%
f_{J}=f_{J0}\left( 1-\eta ^{2}\right) ^{1/4}$, disappearing altogether at a
bias equal to the junction critical current.

\subsection{Large amplitude oscillations}

For larger oscillation amplitudes, the well shape becomes increasingly
anharmonic and oscillations will occur at a frequencies slightly lower than
the small amplitude values given by $f_{J}=f_{J0}\left( 1-\eta ^{2}\right)
^{1/4}$. \ This can be expressed approximately as

\begin{equation}
f_{J}=f_{J0}\sqrt{\;\left[ J_{0}(a)+J_{2}(a)\right] \sqrt{1-\left( \frac{%
\eta }{J_{0}(a)}\right) ^{2}}}  \label{Eq12}
\end{equation}%
where $J_{0}$ and $J_{2}$ are Bessel functions of the first kind and $a$ is
the amplitude of the oscillation. A `large' amplitude oscillation would take
the phase from the minimum position of a well to the inflection point of
that well. In this case the amplitude can be approximated by%
\begin{equation}
a\approx \sqrt{\frac{4}{3}\left( 1-\eta \right) }  \label{a}
\end{equation}%
(see \cite{NGJ2}). At the critical bias $\eta =1$, $a=0$ and $f_{J}=0$.\ For
bias currents less than unity, the anharmonic plasma frequency is slightly
smaller than the harmonic value.

\subsection{Escape from a well}

\begin{figure}[t]
\begin{center}
\scalebox{0.25}{\centering \includegraphics{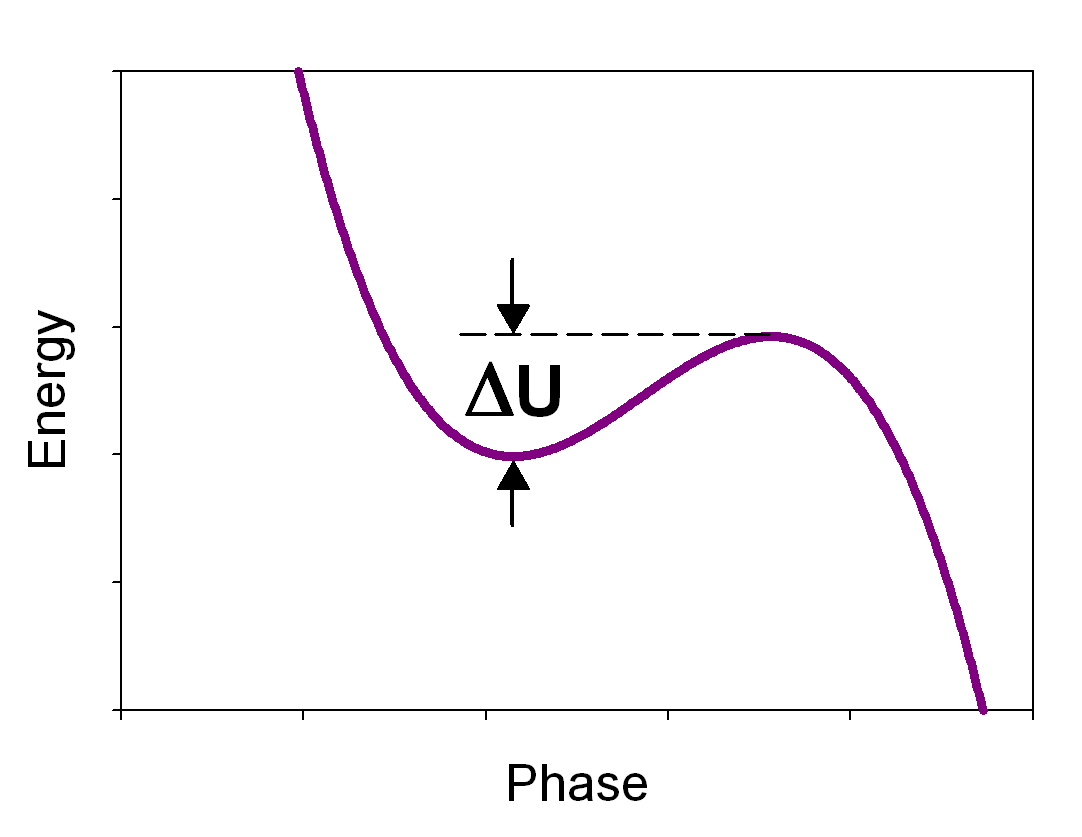}}
\end{center}
\vspace{-0.5 cm}
\caption{Portion of washboard potential
illustrating the barrier for escape.}
\label{Fig2}
\end{figure}

The barrier height $\Delta U$ for the washboard potential, indicated in Fig %
\ref{Fig2}, is controlled by the bias current $\eta $ and is given by Eq.\ref%
{Eq5}

\begin{equation}
\Delta U=2E_{J}\left[ \sqrt{\left( 1-\eta ^{2}\right) }-\eta \cos ^{-1}\eta %
\right]  \label{Eq5}
\end{equation}

If noise is present, then the `particle' can hop out of the well and bounce
down the washboard, as illustrated above. For thermal activation, the escape
rate at temperature $T$ is given by

\begin{equation}
\Gamma (t)=f\exp \left( \frac{-\Delta U}{k_{B}T}\right)  \label{Eq6}
\end{equation}

where $f$ is an attempt frequency. So if a constant bias is applied, escape
is inevitable, but for a shallow well it will happen sooner than for a deep
well.

Because $d\varphi/dt=2eV/\hbar$, this bouncing motion is accompanied by an
oscillating but always positive junction voltage.

Modeling some types of experiment, those involving junctions in
superconducting circuits, can be done by determining the junction phase
dynamics from Eq. \ref{Eq1} while other types, those based on activation
processes, can be addressed with Eq. \ref{Eq2}.

\subsection{Pendulum Analog}

\begin{figure}[t]
\begin{center}
\scalebox{0.25}{\centering \includegraphics{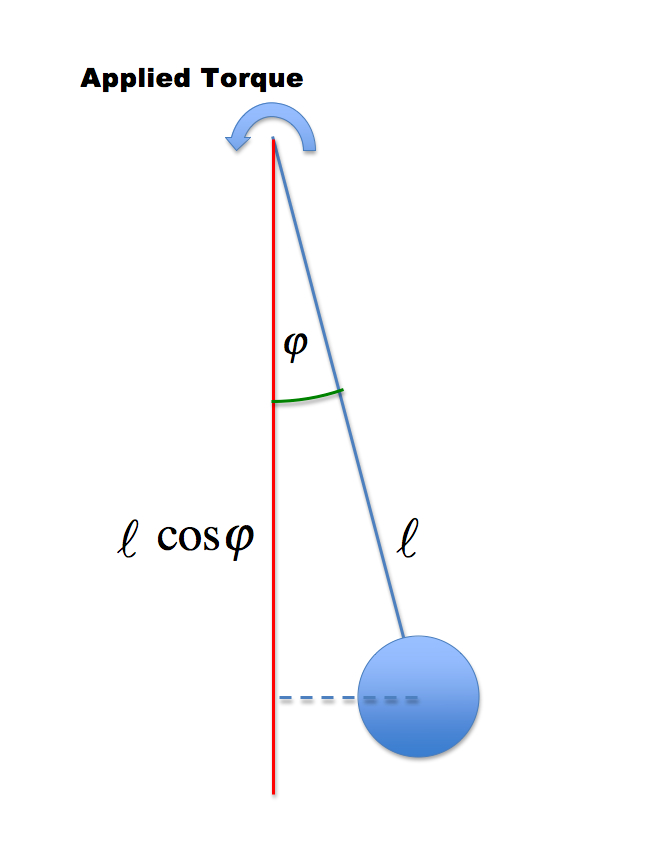}}
\end{center}
\vspace{-0.5 cm}
\caption{A simple pendulum.}
\label{simplependulum}
\end{figure}

Equation \ref{Eq1} is isomorphic to the equation that governs the motion of
a driven pendulum, as depicted in Fig.\ref{simplependulum}.

For the pendulum, the equation of motion is%
\begin{equation}
I\,\ddot{\varphi}+\gamma \dot{\varphi}+mg\ell \sin \varphi =\Gamma
\label{pendulum}
\end{equation}%
where $m$ is the mass of the bob, $\ell $ is the length of the suspension, $%
I $ is the rotational moment of inertia ($m\ell ^{2}$ in the case of a
simple pendulum), $\gamma $ is the damping, and $\Gamma $ is the net appied
torque. \ The potential energy of this system for a displacement angle $%
\varphi $ and an applied torque $\Gamma $ is%
\begin{equation}
U=mg\ell (1-\cos \varphi )+\Gamma \varphi  \label{Pendulum energy}
\end{equation}%
This, of course, is the washboard potential of Eq.\ref{Eq4} and Fig.\ref%
{Fig1}. \ A mechanical analog such as this helps to visualize the behavior
of a Josephson junction. \ For example, increasing the applied torque will
increase the pendulum angle until it reaches $\pi /2$ at which point a
minute additional applied twist, or noise, will cause the pendulum to flip
over the top and then spin - i.e. bounce down the washboard. \ This critical
torque, $\Gamma _{C}=mg\ell $, plays the role of the critical current $I_{C}$
in the junction.

Also, at zero-torque the frequency of small oscillations about the down
position is the familiar $f_{0}=\frac{1}{2\pi }\sqrt{g/\ell }$. which is the
analog of the Josephson zero-bias plasma frequency with the correspondence: $%
\hslash C/2e\Leftrightarrow m\ell ^{2}$. But if these oscillations have a
large motion side-to-side with an amplitude $\theta $ then the period becomes

\begin{equation}
P=4\sqrt{\frac{\ell }{g}}\int_{0}^{\pi /2}\frac{d\theta }{\sqrt{1-k^{2}\sin
^{2}(\theta )}}  \label{period}
\end{equation}%
where the right hand side is an elliptic integral of the first kind (see 
\cite{elliptic}). This period becomes larger than the small angle value as
the amplitude of the oscillations increases, and so the anharmonic frequency
diminishes accordingly.

The analog of a constant current bias of a Josephson junction is a constant
torque applied to the pendulum. For any applied torque $0<\Gamma <\Gamma
_{C} $, the pendulum will have a stable position at an angle somewhere
between $0$ and $\pi /2$. \ Any perturbation will cause oscillations around
that stable point. \ As with the Josephson junction, the frequency of small
oscillations would scale with the applied torque.%
\begin{equation}
f_{P}=f_{0}\sqrt[4]{1-\left( \Gamma /\Gamma _{C}\right) ^{2}}
\label{pendfrequency}
\end{equation}%
But for larger induced oscillations around the stable angle, an expression
similar to Eq.\ref{Eq12} would apply.

\subsection{Switching Current Distributions}

The usual experimental approach to measuring activation involves \textit{%
sweeping} the bias current upward from zero. Early in the sweep, the well is
deep; late in the sweep the well becomes quite shallow. Thermal activation
is easier from a shallow well. But in addition, as the bias is swept, the
plasma frequency drops and thus the rate of escape attempts decreases. The
interplay of these two factors determines the net escape rate for the
momentary bias current.

When the bias current is \textit{repeatedly} ramped up from zero, at a given
temperature, for each sweep the value of the bias current when the junction
switches to the finite voltage state is recorded. The distribution of these
accumulated escape events forms a Switching Current Distribution (SCD) peak
whose two key attributes are position and width; this is the primary form of
experimental data.

A method for simulating the escape process seen in an experiment was
described in \cite{Blackburn1}. The method begins with an ensemble of $M$
junctions. The bias on all junctions starts at $0$ and is incremented in $N$
steps, with each step of duration $\Delta t=\left( Nf_{S}\right) ^{-1}$,
where $f_{S}$ is the sweep frequency. Each step is assigned a channel, and
the total counts in that channel indicates how many junctions have switched
to the finite-voltage state (escaped from a potential well) during that
interval. As the bias sweep proceeds, the original ensemble will have lost $%
e_{1}$ junctions in the first interval, $e_{2}$ junctions in the second
interval, and so forth. Consequently, at the beginning of the $n^{th}$ bias
interval, there will be $M-\sum_{j=1}^{n-1}e_{j}$ junctions not yet escaped.
The number from this remaining pool of junctions that will escape during the
next interval$\Delta t$ will be

\begin{equation}
e_{n}=\left[ M-\sum_{j=1}^{n-1}e_{j}\right] \,\Gamma (t_{n})\Delta t,\quad
n=2,3\cdots N  \label{Eq7}
\end{equation}%
where $\Gamma (t_{n})$ is the probability of escape per unit time in the $%
n^{th}$ interval. Of course, the initial interval just satisfies:

\begin{equation}
e_{1}=M\Gamma (t_{1})\Delta t  \label{Eq8}
\end{equation}

These two equations will mimic a swept-bias experiment provided a suitable
expression is available for the escape rate $\Gamma$.

For the case of thermal activation (TA) out of the potential well at the $%
n^{th} $stage, the escape rate is

\begin{equation}
\Gamma (t_{n})=f_{n}\exp \left( \frac{-\Delta U_{n}}{k_{B}T}\right)
\label{Eq9}
\end{equation}%
\ where $f_{n}$ is the Josephson plasma frequency in the $n^{th}$ bias
interval and $\Delta U_{n}/k_{B}T$ is the height of the potential barrier
divided by the mean thermal energy.

So from Eq. \ref{Eq2} with $\eta _{n}$ indicating the bias current at the $%
n^{th}$ step normalized to the junction critical current, 
\begin{equation}
\Delta U_{n}=2E_{J}\left[ \sqrt{1-\eta _{n}^{2}}-\eta _{n}\cos ^{-1}\eta _{n}%
\right]  \label{Eq10}
\end{equation}

To perform a simulation, only three parameters must be known: the critical
current of the junction $I_{C}$, the bias sweep frequency $f_{S}$, and the
temperature $T$. Values must be chosen for the number of channels $N$ and
for the ensemble number $M$ ; these are arbitrary -- large numbers improve
precision but slow the computation slightly. Typical choices were: $%
N=5000\;\,M=100,000$.

It is worth noting that a number of papers reported on successful
investigations of potentials of specific Josephson junctions systems by
using thermal escape techniques in parallel with RCSJ simulations. These
analyses were concerned with extended Josephson junctions \cite{Kautz},
dc-SQUID systems \cite{NGJ0} and step-edge junctions \cite{Barb} of high Tc
superconductors.

\section{\protect\large The Classical to Quantum Crossover}

The concept of crossover temperature was introduced by Affleck \cite{Affleck}
in the attempt to distinguish between classical and quantum fluctuations in
one-dimensional potentials \ In terms of Josephson quantum narrative, in the
vicinity of the crossover temperature, the mechanism for escape to a finite
voltage state changes from thermal activation (TA) to macroscopic quantum
tunneling (MQT). Because the MQT tunneling rate is \textbf{not} temperature
dependent, the position and width of SCD peaks must both saturate at low
temperatures.

As discussed earlier, the classical escape rate for thermal activation is $%
\Gamma =f_{J}\exp \left( -\frac{\Delta U}{k_{B}T}\right) $ . For the case of
escape via macroscopic quantum tunneling, the rate is given by (see e.g., 
\cite{Devoret}).

\begin{equation}
\begin{array}{l}
{\Gamma_{q}=a_{q}\;f_{J}\exp\left[ -7.2\frac{\Delta U}{hf_{J}}\left( 1+\frac{%
0.87}{Q}\right) \right] } \\ 
{a_{q}\approx\left[ 120\pi\left( \frac{7.2\Delta U}{hf_{J}}\right) \right]
^{1/2}}%
\end{array}
\label{Eq18}
\end{equation}

The crossover temperature may be defined by the condition that the escape
rate for thermal activation equals the escape rate for MQT, which is%
\begin{equation}
f_{J}\exp \left( \frac{-\Delta U}{k_{B}T}\right) =a_{q}\;f_{J}\exp \left[
-7.2\frac{\Delta U}{hf_{J}}\left( 1+\frac{0.87}{Q}\right) \right]
\label{Eq19}
\end{equation}

and so

\begin{equation}
\frac{\Delta U}{k_{B}T}\approx7.2\frac{\Delta U}{hf_{J}}  \label{Eq20}
\end{equation}
with the usual assumptions: $a_{q}\approx1$ and$Q\gg1$. Therefore

\begin{equation}
T_{cr}=\frac{hf_{J}}{7.2k_{B}}  \label{Eq21}
\end{equation}%
The Josephson plasma frequency is bias-dependent because the curvature of
the well is also bias-dependent. For example, in the harmonic approximation, 
$\ f_{J}=f_{J0}\left( 1-\eta ^{2}\right) ^{1/4}$.

\begin{figure}[t]
\begin{center}
\scalebox{0.25}{\centering \includegraphics{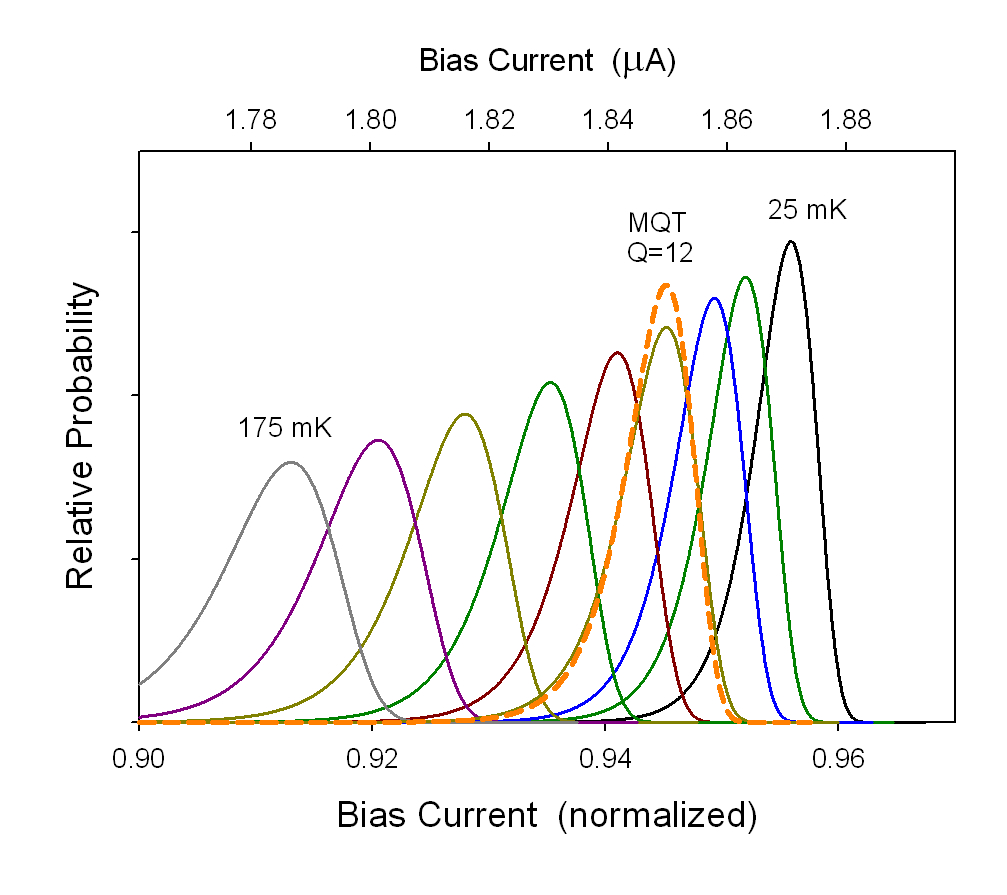}}
\end{center}
\vspace{-0.5 cm}
\caption{Results of two simulations using
the parameters of the sample in \protect\cite{Yu}. One simulation utilized
the thermal activation (TA) escape rate, while the other used the
macroscopic tunneling (MQT) escape rate, Eq.\protect\ref{Eq18}. The MQT peak
coincides with the particular TA peak for $65$ \textit{mK.}}
\label{Fig23}
\end{figure}

Switching current distributions (SCD) are acquired by repeatedly sweeping
the bias current at a specified temperature; consequently throughout any
scan the TA escape rate varies only because of the changing bias. For each
temperature, there will be a different SCD peak. In comparison, because the
MQT escape rate is independent of temperature, only a single peak should be
seen. Figure \ref{Fig23} shows simulation results based on experimental data
from \cite{Yu}.

It can be seen that the MQT SCD peak
falls exactly on a TA SCD peak at just one particular temperature -- that
defines the crossover temperature. Note that the value of the bias of this
special peak is not zero, and so the plasma frequency $f_{J}$ will be
smaller than $f_{J0}$. Because any peak position is never known \textit{a
priori}, the expression for the crossover temperature can only be used as an
approximation with $f_{J}\approx f_{J0}$. Suppose for example, an
experimental SCSD peak happened to occur at a normalized bias of $0.9$. \
Then $\left( 1-0.9^{2}\right) ^{1/4}=0.66$ and so the anticipated crossover
temperature would be a fraction of the value obtained using $\ f_{Jo}$.
Taking as an example a crossover temperature quoted at $60$ \textit{mK}, the
`true' value could be $40$ \textit{mK}. The higher the peak position, the
larger the discrepancy. Most cryostats bottom out at around $20 $ \textit{mK}
meaning the acquired data do not reach very far below the presumed crossover
value. We conclude that there is not much room to show unambiguous
experimental confirmation of an MQT transition in Josephson junctions.

\subsection{Experimental Evidence}

\subparagraph{Voss and Webb}

\begin{figure}[t]
\begin{center}
\scalebox{0.5}{\centering \includegraphics{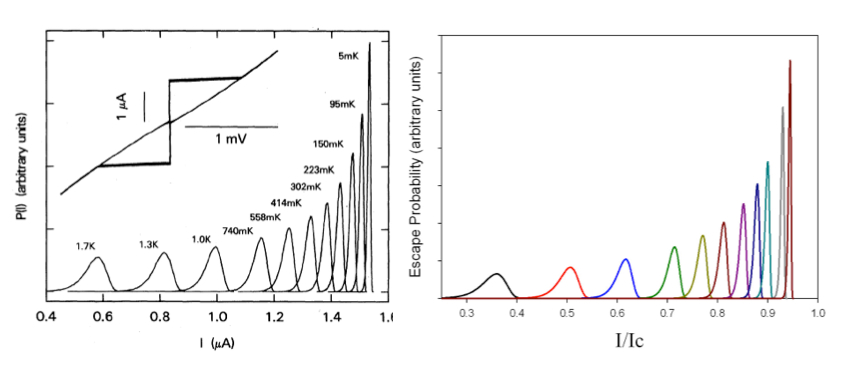}}
\end{center}
\vspace{-0.5 cm}
\caption{Experimental results
from \protect\cite{VossWebb} for SCD peaks at various temperatures; Right: RCSJ
simulation results}
\label{Fig3}
\end{figure}

The pivotal moment in the history of switching current distributions
measured at millikelvin temperatures was the 1981 paper by Voss and Webb 
\cite{VossWebb}. They presented experimentally recorded SCD peaks for 11
selected temperatures. Their figure is shown here (Fig \ref{Fig3}) together
with results from an RCSJ simulation based on the thermal-activation escape
rate given in Eq.\ref{Eq9}.

The simulation is clearly in excellent agreement with experiment, but a
precise test is not possible because full details of the bias sweep were not
given; the text says only that \textquotedblleft A low-frequency sinusoidal
current (amplitude $\approx 2I_{C}$) was applied to the junction and the
distribution of currents at which the junction switched out of the
superconducting state was measured\textquotedblright . The current-sweeps
were stipulated as being in the range $10-20\,Hz$. Peak positions,
especially at high bias values, are known to be sensitive to the shape and
speed of the bias sweep.

\begin{figure}[t]
\begin{center}
\scalebox{0.25}{\centering \includegraphics{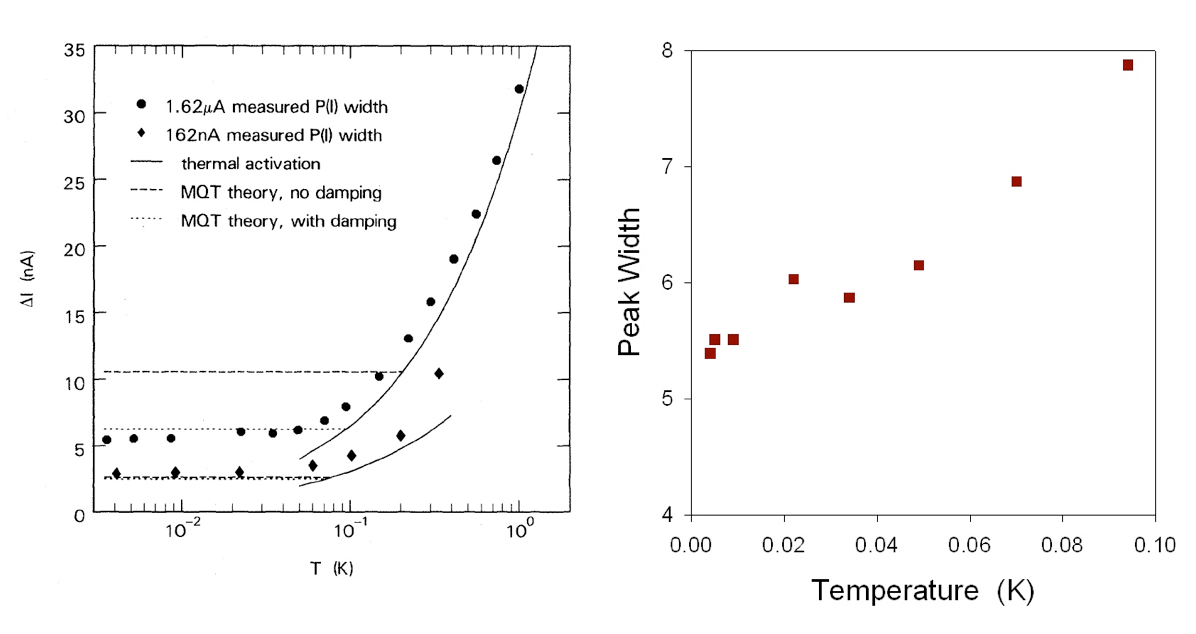}}
\end{center}
\vspace{-0.5 cm}
\caption{Left:
Experimental data of \protect\cite{VossWebb}; Right: Expanded view of data on the
left at the lowest temperatures and with a linear temperature scale.}
\label{Fig24}
\end{figure}

The peak width data from Voss and Webb \cite{VossWebb} for the sample with
critical current $1.62{\;}\mu A$ are shown in Fig.\ref{Fig24}. Note the
choice of a log scale for temperature. Again, this set the standard for
presentations of experimental data (however see the Appendix). From this
figure data points were extracted and then re-plotted as shown in the right
panel.

This illustrates a crucial point: the choice of a logarithmic temperature
axis makes it appear that there is a flattening. The linear plot of the same
data, shown here, makes it very clear that saturation of the peak width
below $100\,mK$ has \textbf{not} occurred. The fact that the trend in the
data is not towards the origin is discussed in the section: Effective
Temperature.

Since the paper by Voss and Webb \cite{VossWebb}, many other swept bias
experiments have been reported. A typical example from three decades later
is the following.

\subparagraph{Yu et al.}

\begin{figure}[t]
\begin{center}
\scalebox{0.5}{\centering \includegraphics{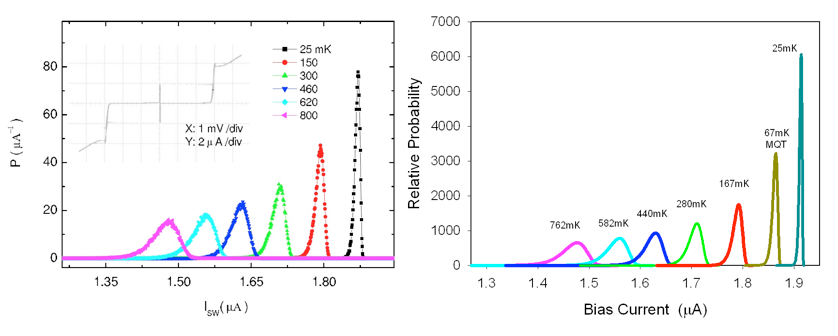}}
\end{center}
\vspace{-0.5 cm}
\caption{Left: Experimental
results for swept bias escape peaks from \protect\cite{Yu}; Right: RCSJ simulation results
for escape peaks.}
\label{Fig4}
\end{figure}

Yu et al. \cite{Yu} carried out swept bias experiments on a junction with a
critical current of $I_{C}=1.957\,\mu A$ and a bias sweep rate of $0.4$
mA/s; the observed SCD peaks are shown in Fig.\ref{Fig4}. An RCSJ simulation
was applied to this experiment \cite{Blackburn3}; representative simulation
SCD peaks are also shown.

\begin{figure}[t]
\begin{center}
\scalebox{0.25}{\centering \includegraphics{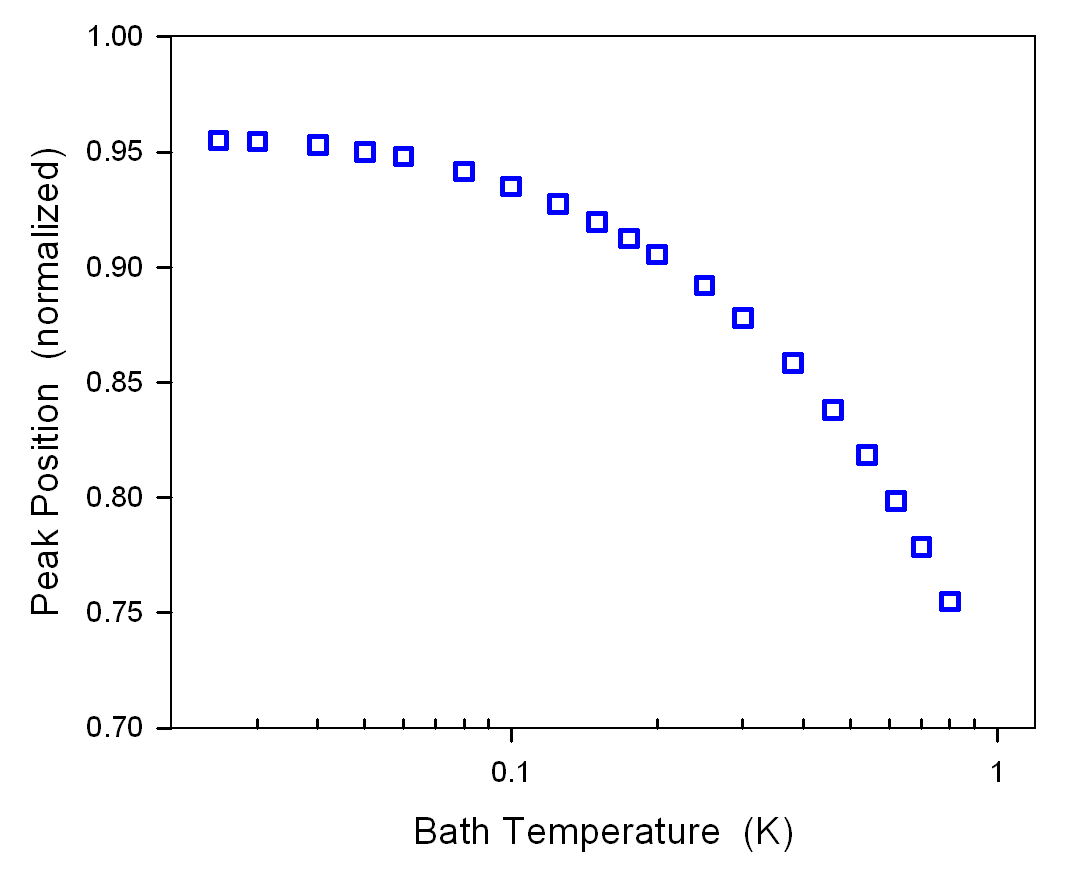}}
\end{center}
\vspace{-0.5 cm}
\caption{SCD peak position data as
presented in Fig.2 of \protect\cite{Yu}.}
\label{YuPeakPositions}
\end{figure}

From figure 2 in \cite{Yu} the experimental peak positions (blue squares)
were digitized and are plotted in Figure \ref{YuPeakPositions}. \ Clearly,
there is a suggestion of levelling off at low temperatures, however this
impression is exaggerated by the choice of a logarithmic temperature scale. 

\subsection{Effective Temperature}

\begin{figure}[t]
\begin{center}
\scalebox{0.5}{\centering \includegraphics{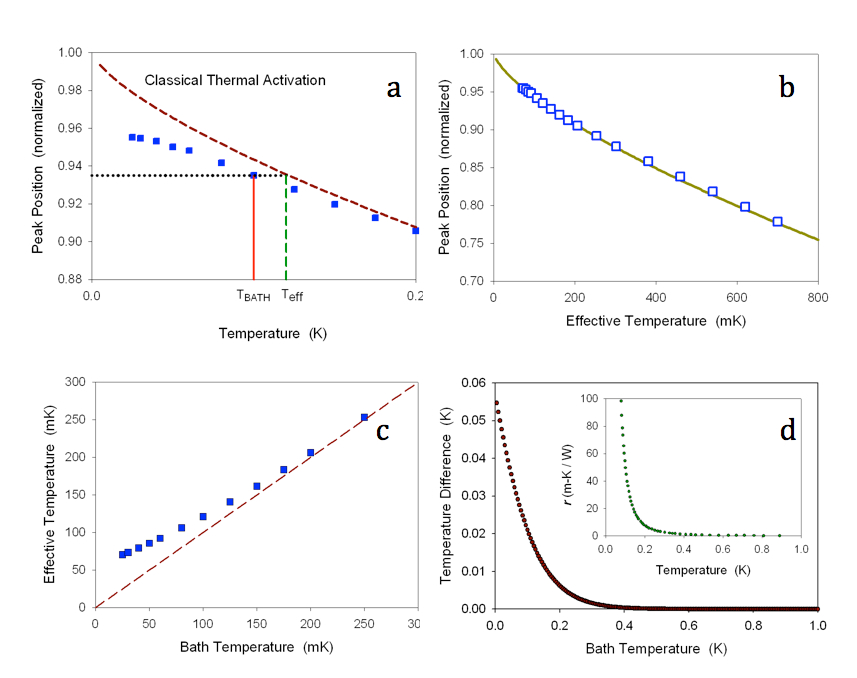}}
\end{center}
\vspace{-0.5 cm}
\caption{(a) Experimental data from the preceding figure (squares)
plotted with a linear temperature scale, and the corresponding results from
a simulation based thermel activation model. The constuction lines indicate
how a corrected sample temperature can be inferred.  (b) Experimentalpeak
positions plotted versus effective sample temperature  (c) Effective sample
temperature versus bath temperature.  (d) Comparison of the temperature
elevation implied by the effective temperature with thermal resistivity data
for niobium (inset)}
\label{correction}
\end{figure}

Just replacing the log scale with a linear scale produces the result shown
in Fig.\ref{correction} (a). Clearly the peak positions deviate from the
expectations of a fully classical TA escape process (dashed line) and the
deviation increases as the temperature is lowered. Nevertheless, it is quite
apparent that the peak positions have not decisively saturated at the lowest
temperatures. A resolution of this difficulty can be found in choosing any
specific peak, and noticing that the experimentally reported temperature is
not the same as the temperature of an identical classical peak, as indicated
by the construction lines in Fig.\ref{correction} (a).

In other words, it is the effective sample temperature that ought to be used
in the escape rate
\begin{equation}
\Gamma (t_{n})=f_{n}\exp \left( \frac{\Delta U_{n}}{k_{B}T_{eff}}\right)
\label{Eq11}
\end{equation}%
to give correct peak positions (and widths). Our definition of $T_{eff}$ is
essentially the same as the escape temperature $T_{esc}$ introduced in 1985
by Devoret, Martinis, and Clarke \cite{Devoret} and repeated in several
subsequent papers. With the bath temperature replaced by the effective
temperature, the earlier plot changes to Fig.\ref{correction}(b) which shows
excellent agreement between experimental data and predictions of thermal
activation theory.

The experimentally reported temperatures are, in reality, the measured
mixing chamber temperatures. The effective temperatures refer to the
junctions themselves. A plot of $T_{eff}\,vs\,T_{BATH}$ (see Fig.\ref%
{correction} (c)) highlights the degree to which the junction temperature
rises above the bath temperature.

Even more revealing is a plot, Fig.\ref{correction} (d), derived from the
previous graph, of the difference between the effective (junction)
temperature and the bath temperature as a function of bath temperature. The
junction temperature is seen to begin rising above the bath temperature
below about $300mK$ and this behavior is mirrored in the temperature
dependence of the thermal resistivity of niobium (inset). One could infer
from these plots that the apparent rise of the junction temperature might be
connected to the increasing thermal resistance of the thermal path from
junction to bath.

The paper by Voss and Webb in 1981 indeed launched a field of study; its
importance was that it seemed to show for the first time (in Fig.3) clear
evidence that \textquotedblleft Below $100${\ }\textit{mK }the distribution
widths become independent of $T$\textquotedblright\ and that
\textquotedblleft The results are in excellent agreement with predictions
for the quantum tunneling of the (macroscopic) junction phase \dots
\textquotedblright , in other words, that MQT had been confirmed.

\begin{figure}[t]
\begin{center}
\scalebox{0.25}{\centering \includegraphics{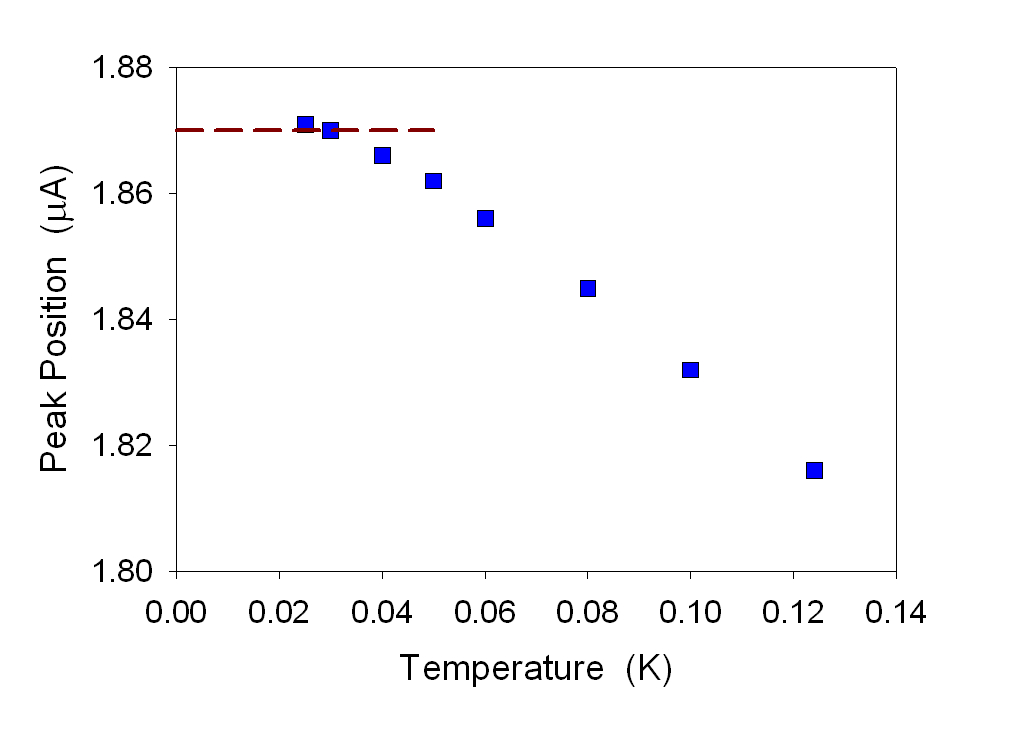}}
\end{center}
\vspace{-0.5 cm}
\caption{Expanded portion of SCD peak
position data from figure 2 in \protect\cite{Yu} but with a linear
temperature scale. The horizontal dashed line is the MQT prediction
according to \protect\cite{Yu}.}
\label{Fig25}
\end{figure}

Figure \ref{Fig25} is an expanded view of experimental data points from Yu
et al.\cite{Yu}. Both axes are linear in contrast to a log temperature scale
which, as was pointed out earlier, tends to stretch the data along the
temperature axis and create the appearance of a saturation of the peak
positions as the lowest temperatures are approached. The dashed line is
drawn at the peak position indicated by Yu et al. as the expected quantum
limit.

It is quite apparent that the peak positions have \textbf{not} become
temperature independent, and no evidence of an MQT transition can be claimed.

A recent investigation of transmon qubits \cite{Jin} considered the
possibility that a superconducting qubit might not be in equilibrium with
its cryogenic environment, leading to effective temperatures in the range $%
50-130mK$.

\subsection{Escape Temperature}

In 1985, Devoret et al. \cite{Devoret} introduced the concept of an escape
temperature, defined by the expression

\begin{equation}
\Gamma =\frac{\omega _{J}}{2\pi }\exp \left( -\frac{\Delta U}{k_{B}T_{esc}}%
\right)  \label{Tesc1}
\end{equation}

From their experimental data for the positions of the escape peaks at a each
bath temperature, the escape rate was determined. Then, $T_{esc}$was
calculated from the above equation. In other words, they asked: what revised
junction temperature would yield agreement with the experimental
observations of escape peak positions? This in fact is exactly the question
posed in our discussion of SCD peaks in the RCSJ model. So the escape
temperature here is the effective temperature in the earlier discussion.

\begin{figure}[t]
\begin{center}
\scalebox{0.25}{\centering \includegraphics{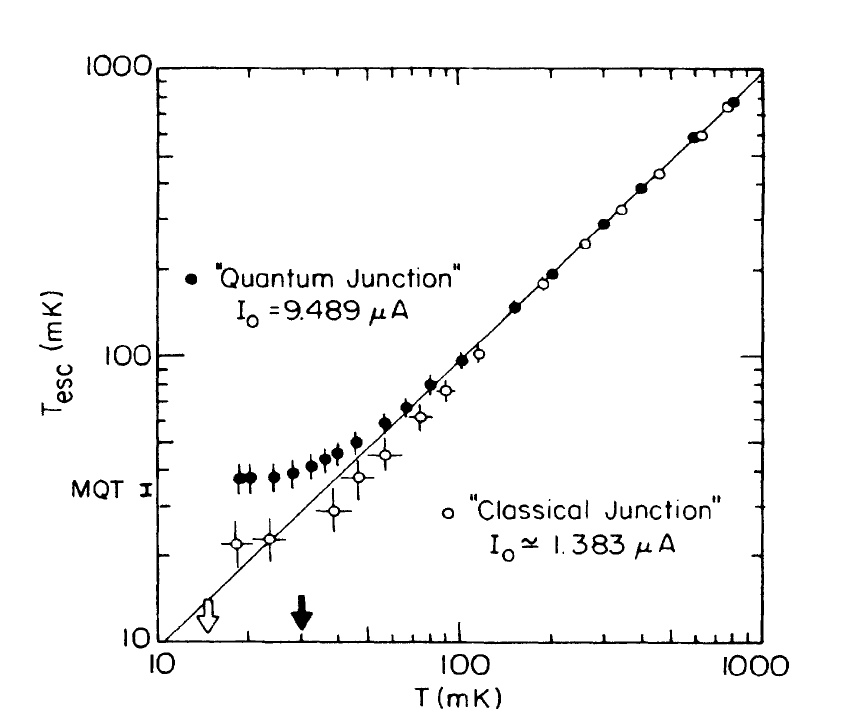}}
\end{center}
\vspace{-0.5 cm}
\caption{Plot of escape temperature versus
bath temperature as reported in \protect\cite{Martinis}.}
\label{Fig26}
\end{figure}

Their plot showing the relationship between escape temperature and bath
temperature is reproduced here in Fig.\ref{Fig26}.

In the limit $T\to0$, the escape rate is expected to be entirely due to MQT
and therefore

\begin{equation}
T_{esc}\rightarrow \frac{\hbar \omega _{J}}{7.2k_{B}\left( 1+0.87/Q\right) }
\label{Tesc2}
\end{equation}

For their high critical current junction, they indicate that the escape
temperature was expected to be $36$ \textit{mK}, as marked by the label
`MQT'.

The calculation can be inverted by substituting this value of $T_{esc} $ in
the expression and asking what plasma frequency is implied. For $Q=30$ the
conclusion is that to make this work, the plasma frequency must be $%
\omega_{J} =35.9\times10^{9} $ or $f_{J} =5.71\, GHz$. However, the
zero-bias plasma frequency $\omega_{J0} =\sqrt{2eI_{C} /\hbar C} $ equals $%
67.4\times 10^{9} $ using the stated values $I_{C} =9.489\mu A$ and $%
C=6.35\, pF$; thus $f_{J0} =10.72\, GHz$. Hence the \textit{apparent} fit to
the experimental data in the $T\to0$ limit is a result of choosing a plasma
frequency approximately half the zero-bias value. That drop in value would
occur at a normalized bias of $0.9375$. But what justification is there for
this value? In Devoret et al. the claim was made that all of this is
achieved ``with no adjustable parameters'', but it seems the plasma
frequency was selected to achieve the desired limit.

An important point to note is that each of the points in this plot came from
finding an escape temperature that matched the observed escape rate $\Gamma$%
. There is nothing intrinsically quantum in such a calculation. Only the
limiting value ($T\to0$) is a prediction of MQT, not the data points at
finite temperature.

This particular plot first appeared in Devoret, Martinis, and Clarke (1985) 
\cite{Devoret}, and the exact same figure appeared also in Martinis,
Devoret, and Clarke (1987) \cite{Devoret} and yet again in Clarke, Cleland,
Devoret, Esteve, and Martinis (1988) \cite{Devoret}. It might also be noted
that the \textquotedblleft Classical Junction\textquotedblright\ was just
the \textquotedblleft Quantum Junction\textquotedblright\ with an applied
magnetic field used to reduce the critical current from $9.489\mu A$ to $%
1.383\mu A$, so in fact only one experimental sample was involved in all
these claims of observing MQT effects.

\begin{figure}[t]
\begin{center}
\scalebox{0.25}{\centering \includegraphics{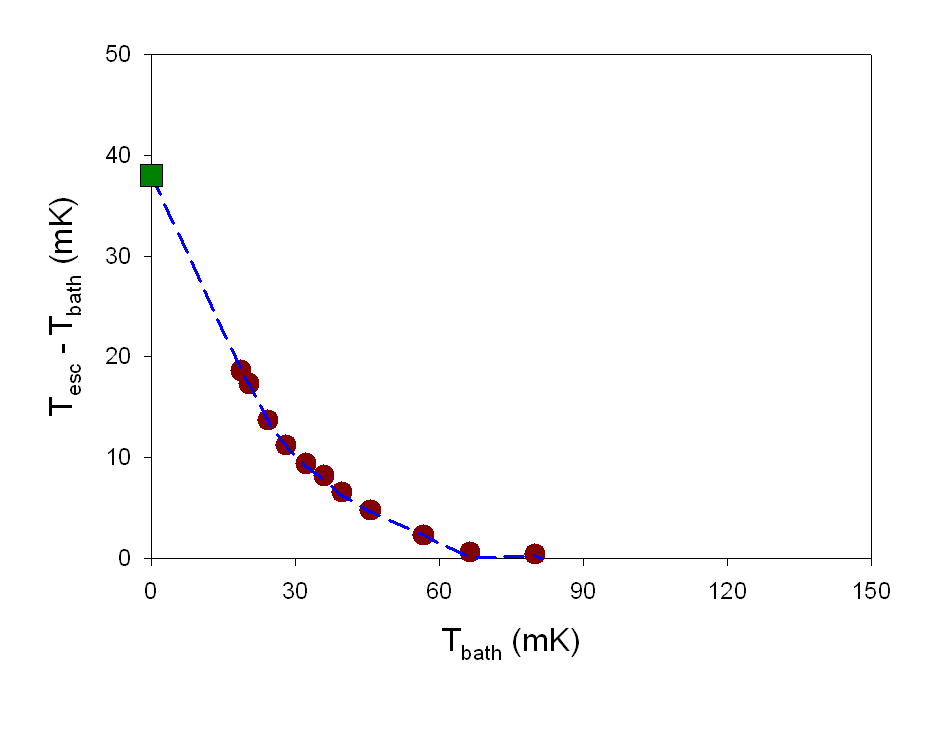}}
\end{center}
\vspace{-0.5 cm}
\caption{Data from Fig.\protect\ref{Fig26}
replotted to show the elevation of sample temperature as a function of bath
temperature.}
\label{Fig27}
\end{figure}

As was done in the earlier discussion of the RCSJ model applied to such
experiments, an alternate graphical presentation (Fig.\ref{Fig27}) hints at
another interpretation.

Taking all the data points and not focusing entirely on the limiting value,
it becomes apparent, as discussed for the RCSJ interpretation of the data of
Yu et al., that the sample temperature appears to rise above the bath
temperature, Martinis et al. (1987) \cite{Devoret} were obviously aware of
this issue when they stated: \textquotedblleft Although the low-temperature
values of $T_{esc}$ plotted in Fig. 2 are in good agreement with the $T=0$
prediction, nevertheless one should demonstrate that the flattening of $%
T_{esc}$ is not due to an unknown, spurious noise source.\textquotedblright

This crucial point was raised by Cristiano and Silvestrini \cite{Cristiano}
who noted that in the presence of external noise, the effective temperature
can be higher than the bath temperature. This matter was debated in an
exchange in Physical Review Letters 1989 \cite{Silvestrini2, Devoret2} . In
a subsequent paper on the topic of escape temperature, Silvestrini \cite%
{Silvestrini} aptly remarked: \textquotedblleft From a theoretical point of
view, $T_{esc}$ is then a good variable to clearly visualize the transition
between the classical and quantum limit, \textbf{if there is any}%
\textquotedblright\ and then \textquotedblleft From an experimental point of
view, on the contrary, $T_{esc}$ is not a direct measured quantity but
strongly correlated to junction parameters which are not known \textit{a
priori}, as the critical current $I_{C}$ , the junction effective resistance
and capacitance.\textquotedblright\ And in fact Devoret et al. (1985) \cite%
{Devoret} did concede that \textquotedblleft the error in the predicted
value of $T_{esc}$ arises predominantly from uncertainties in $\omega _{P}$
and $Q$ \textquotedblright .

A definitive resolution of the dispute would be to directly measure the
junction temperature during the experiment instead of measuring the bath
temperature. Then one would know for certain if the escape temperature did
or did not indicate an MQT escape process at work. Otherwise, as is obvious,
there remains a `loophole' in the logic because evidence that no noise is
present is indirect and possibly not applicable to the many other similar
experiments that also claim to see MQT.

Finally, in their Reply to Silvestrini's Comment, Devoret et al.\cite%
{Devoret} conceded that: \textquotedblleft It is of course true that, as
Silvestrini states, our data \textquotedblleft do not represent an
unambiguous proof of MQT\textquotedblright .

\section{\protect\large Classical Resonances and Quantum Levels}

\subsection{Classical Resonances}

In terms of RCSJ model the phase dynamics of a Josephson junction are
governed by equation \ref{Eq1}. For small amplitude oscillations in the
washboard potential well, the natural frequency is $f_{J}=f_{J0}\left(
1-\eta ^{2}\right) ^{1/4}$. In the anharmonic approximation, larger
amplitude oscillations occur at a frequency given by Eq.\ref{Eq12}.

When both dc and ac bias currents are applied, the phase equation becomes

\begin{equation}
\ddot{\varphi}+\alpha \dot{\varphi}+\sin \varphi =\eta +\eta _{ac}\sin
\left( \omega t\right)  \label{phase_eq}
\end{equation}

\begin{figure}[t]
\begin{center}
\scalebox{0.25}{\centering \includegraphics{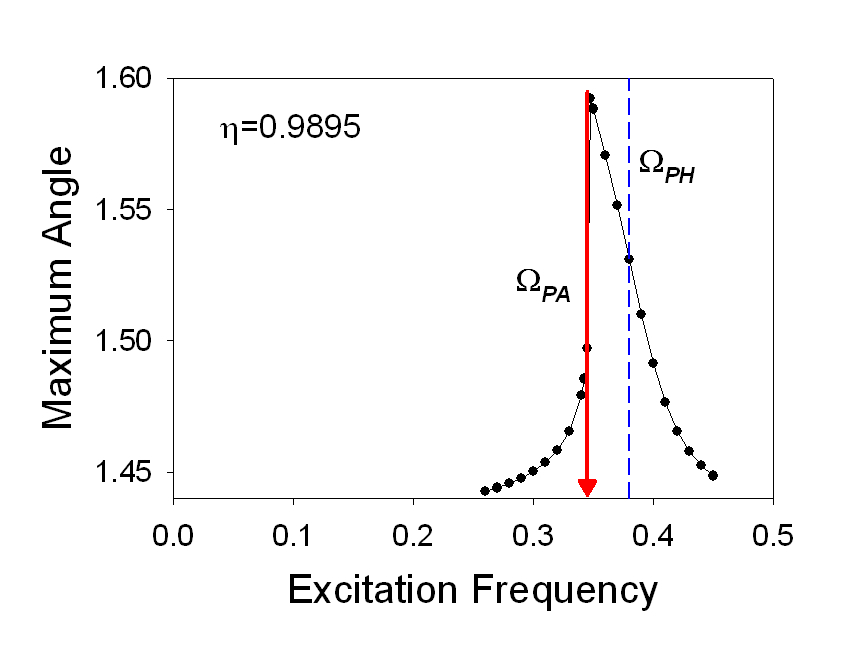}}
\end{center}
\vspace{-0.5 cm}
\caption{Maximum amplitude, on the
barrier side of the well, of induced phase oscillations as a function of the
frequency of excitation. Dots: numerical simulation, $\Omega _{PH}$ is the
plasma frequency in the harmonic approximation, $\Omega _{PA}$ is the plasma
frequency in the anharmonic approximation.}
\label{resonance}
\end{figure}

The ac forcing term mimics the presence of external microwaves irradiating
the junction. As an illustrative example, the following junction parameters
were chosen: $I_{C}=9.489\mu A$,$C=6.35pF$, giving a zero-bias plasma
frequency of $10.72\,GHz$ The above was solved numerically \cite{Blackburn2}%
, see Fig.\ref{resonance}, under the following conditions: the dissipation
constant was set at $\alpha =0.00845$ and the dc bias current was fixed, in
this case at $0.9895$. For each selected ac bias frequency, the excitation
amplitude $\eta _{ac}$ was ramped up from zero to approximately $0.001$ over
a time interval of $600$ plasma periods. The maximum amplitude of the
resulting oscillations in $\varphi $ on the escape side of the well was
noted.

This response curve shows clearly that the amplitude of the induced phase
oscillations is a maximum when the excitation frequency matches the plasma
frequency as determined by the anharmonic approximation (marked by a black
triangle). The plasma frequency calculated from the harmonic approximation
(dashed line) lies slightly above that value and has slightly smaller
induced oscillations.

To carry out a full simulation of a swept bias experiment, the previous
method (no microwaves) must be adapted to the present case. The essential
idea is to modify the previous escape rate (at the $n^{th}$ step in the bias
sweep)

\begin{equation}
\Gamma(t_{n})=f_{n}\exp\left( \frac{-\Delta U_{n}}{k_{B}T}\right)
\label{Eq13}
\end{equation}

so as to account for the lowering of the barrier by the resonant excitation.
Whenever sustained phase oscillations are induced, energy is pumped into the
system.

As figure \ref{Fig14} suggests, the maximum induced phase excursion $\delta
\varphi _{n}$ on the escape side of the well that exists at the $n^{th\text{ 
}}$bias step leads to an effective barrier that is smaller than the full
height $\Delta U_{n}$:

\begin{figure}[t]
\begin{center}
\scalebox{0.25}{\centering \includegraphics{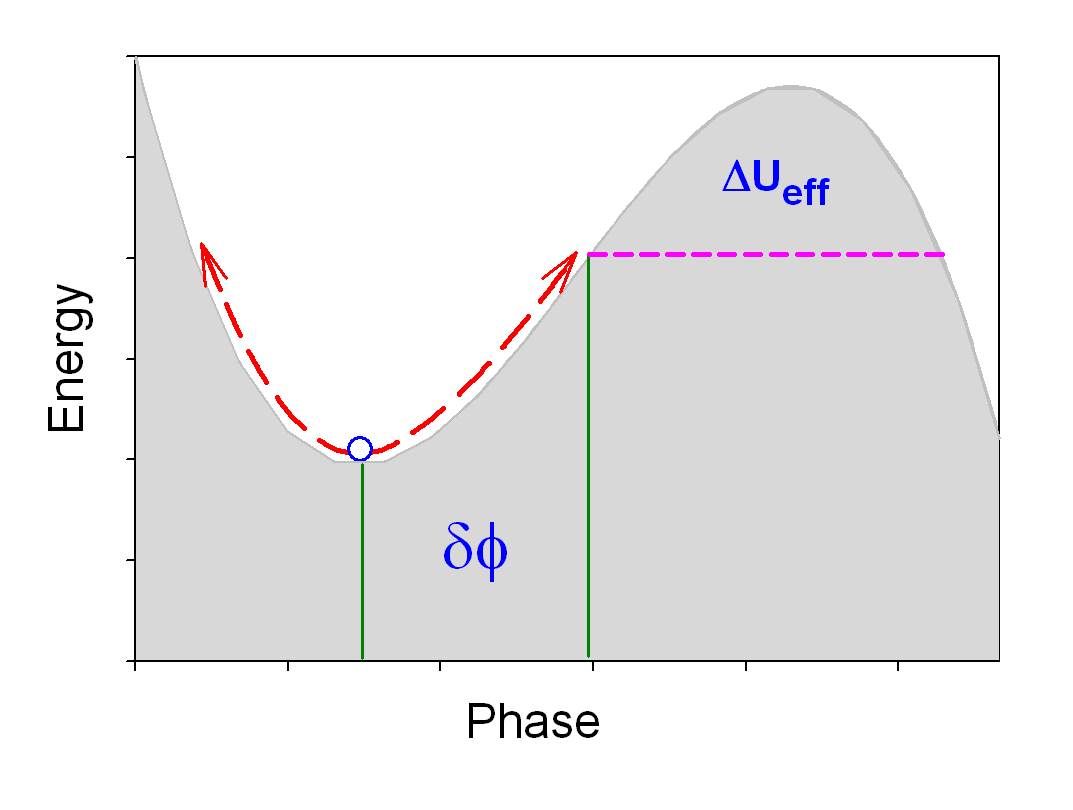}}
\end{center}
\vspace{-0.5 cm}
\caption{Induced oscillations of a
"particle" in the washboard potential well. \ The reduced escape barrier $%
\Delta U_{eff}$ \ is indicated.}
\label{Fig14}
\end{figure}

\begin{equation}
\frac{\Delta U_{eff}}{k_{B}T}=\frac{\Delta U_{n}}{k_{B}T}-\frac{\Delta
U_{pump}}{k_{B}T}  \label{Eq14}
\end{equation}

It is reasonable to suppose that ac induced phase oscillations would take
the form

\begin{equation}
\delta\varphi_{n}=a\frac{b^{2}}{\left( \eta_{n}-\eta_{res}\right) ^{2}+b^{2}}%
,  \label{Eq15}
\end{equation}

In other words, a Lorentzian distribution centered on the resonance
mentioned earlier -- $\eta _{res}$ being the anharmonic bias value.

The pump energy can be calculated from the value of the phase at the bottom
of the well together with $\delta \varphi _{n}$. With this modification to
the barrier height, the revised escape rate can be used exactly as in the
earlier simulation of swept bias experiments without microwaves.

\subsection{Quantum Levels}

\begin{figure}[t]
\begin{center}
\scalebox{0.20}{\centering \includegraphics{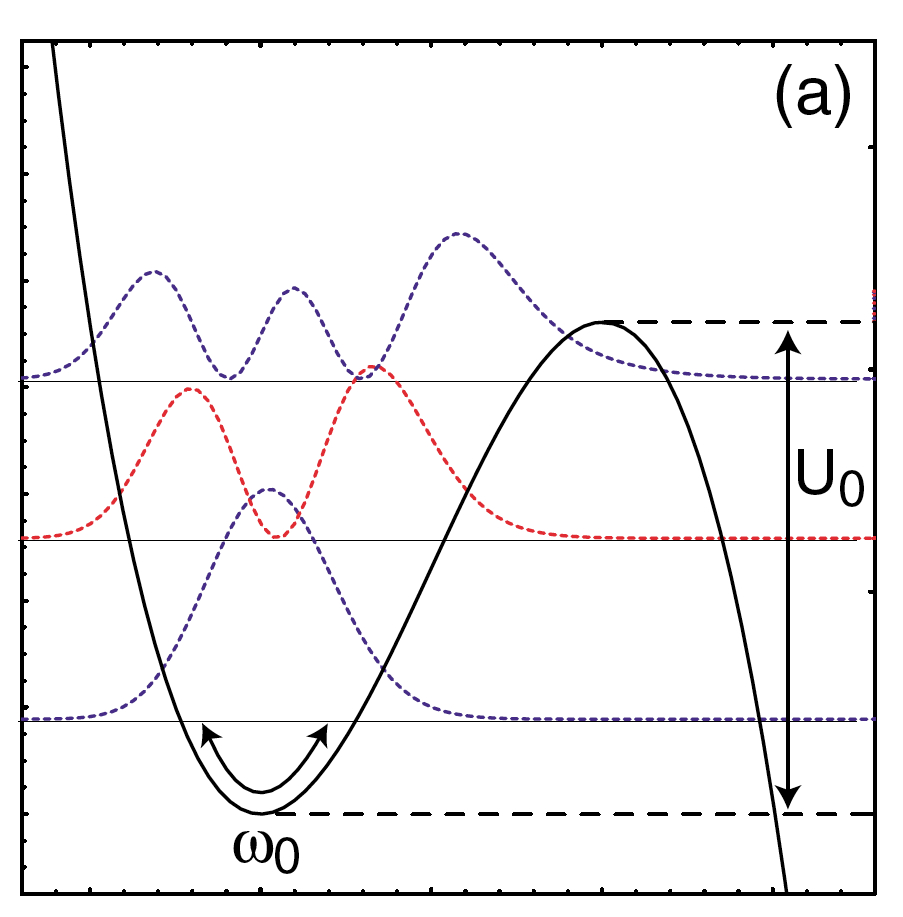}}
\end{center}
\vspace{-0.5 cm}
\caption{First three quantum
wavefunctions in the potential of a Josephson junction (from \protect\cite%
{Wallraff2}).\ The solid curve is $U(\protect\varphi )$ versus $\protect%
\varphi $. The barrier height is $U_{0}$ and the zero-bias plasma frequency
is $\protect\omega _{0}$.}
\label{QuantumLevels}
\end{figure}

The potential energy of a Josephson junction $U(\varphi )$ is given by Eq.%
\ref{Eq4} and is plotted in Fig.\ref{QuantumLevels}. The macroscopic quantum
conjecture is that well below the crossover temperature the phase of a
Josephson junction $\varphi $ becomes the coordinate of a quantum `particle'
existing within a potential well which can contain discrete energy levels.
The wavefunctions of the ground state and first two excited states,
calculated numerically \cite{Wallraff2}, are shown. \ If \ true, a Josephson
junction would be a macroscopic object with atom-like energy states \cite%
{You}. 
The potential $U(\varphi )$ is
scaled by the Josephson energy $E_{J}$ which in turn depends on the junction
critical current $I_{C}$. As indicated in Fig.\ref{Solymar}, below a few
hundred millikelvin the critical current essentially has no temperature
dependence, which means the well retains its shape and scaling at these low
temperatures. One way of viewing the quantum hypothesis would then be to say
that above some \textquotedblleft crossover temperature\textquotedblright , $%
T_{cr}$, there are no wavefunctions or levels, while substantially below the
crossover temperature there are wavefunctions and levels. In other words, if
one imagined a sample with fixed bias $I<I_{C}$, then alternately decreasing
the temperature to perhaps $25$ mK and then increasing the temperature to $%
300$ mK would result in levels emerging and vanishing within the well, with
the eigenfunctions remaining unchanged. This crossover phenomenon would
require new physics \cite{Leggett}- a classical-to-quantum transition.

In elementary quantum theory, tunneling from potential wells is a
temperature independent process. \ This property should distinguish between
classical thermal escape from a well and quantum tunneling out of a well. \
Experimental switching current distributions essentially probe the escape
process. \ Therefore a hallmark of the appearance of a macroscopic quantum
state would be evidence of transitions between the levels, such as $%
\left\vert 0\right\rangle \rightarrow \left\vert 1\right\rangle $, $%
\left\vert 1\right\rangle \rightarrow \left\vert 2\right\rangle $, $%
\left\vert 2\right\rangle \rightarrow \left\vert 3\right\rangle $, $%
\left\vert 3\right\rangle \rightarrow \left\vert 4\right\rangle $, etc. \
Either or both of the first two ac excitation peaks might be confused with
classical ac resonant excitation - the topic of this section.

Evidence of quantum levels in single Josephson junctions has been sought
essentially in two ways, by microwave spectroscopy \cite{Wallraff2,
Martinis, Berkley, Thrailkill} and by very fast sweep of the current-voltage
characteristics \cite{Silve97}. We shall analyze first the microwave
spectroscopy data.

\subsection{Small Junctions}

For the most part, experiments with microwaves are concerned with the 
\textit{position} of escape peaks when junctions are irradiated with
microwaves of fixed frequency; the shape of the resonances plays little role
in the discussion.

\subparagraph{Martinis et al. (1985)}

\begin{figure}[t]
\begin{center}
\scalebox{0.5}{\centering \includegraphics{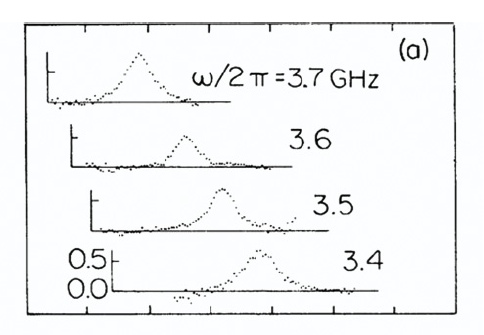}}
\end{center}
\vspace{-0.5 cm}
\caption{Experimental observations of SCD
peaks for different frequencies of applied microwaves.}
\label{Fig12}
\end{figure}

Fig.\ref{Fig12}, reproduced here from figure 3 in \cite{Martinis}, was the
first to show SCD escape peaks in an experiment carried out by applying a
fixed frequency microwave input to the Josephson junction while its dc bias
current is swept. The horizontal axis runs from $9.37\mu A$ to $9.44\mu A$

\begin{figure}[t]
\begin{center}
\scalebox{0.25}{\centering \includegraphics{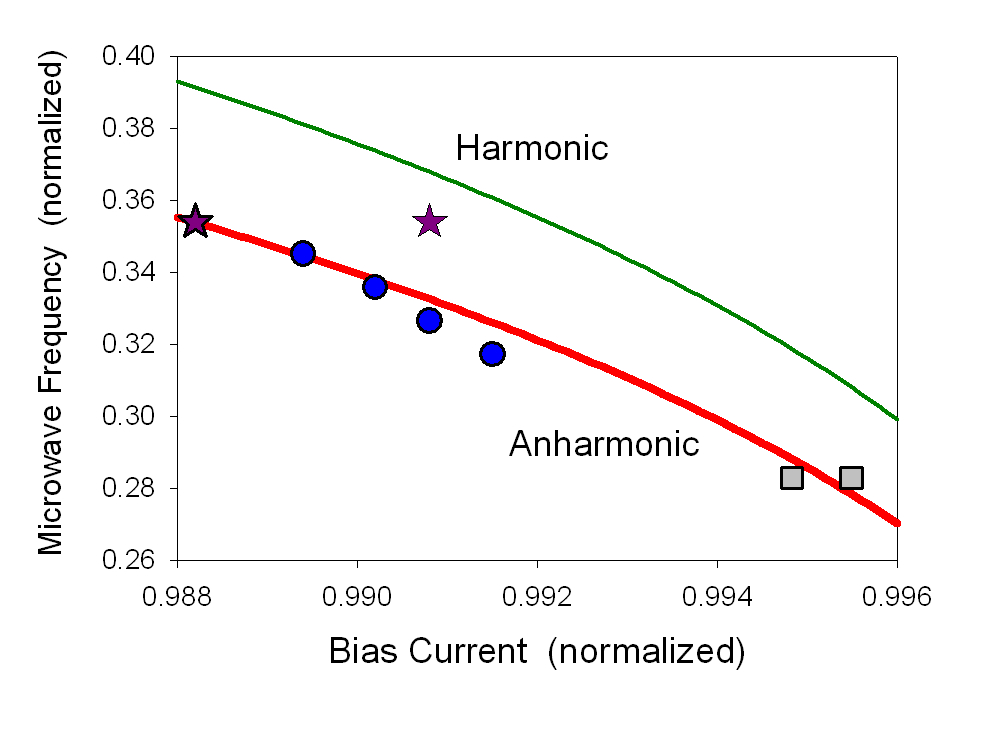}}
\end{center}
\vspace{-0.5 cm}
\caption{Summary of experimental results
from two groups on observed SCD peak positions - circles and squares 
\protect\cite{Martinis}, and stars \protect\cite{Berkley}. \ Continuous
curves were generated from expressions for harmonic and anharmonic resonances.}
\label{Fig13}
\end{figure}

From this figure, peak positions can be extracted and plotted as functions
of the microwave frequency (circles) -- as shown in Fig.\ref{Fig13}. This
has been done as well for two of the peaks in Fig. 2 of \cite{Martinis}
(squares) and, finally, (stars) for the two peaks in Fig. 3 of \cite{Berkley}
(see \cite{Blackburn1, Blackburn2}).

Clearly, the positions of these various experimentally observed peaks
closely match the expectations of anharmonic resonances in the washboard
potential well -- a prediction of the RCSJ model. In the case of two peaks
seen by Berkley et al. \cite{Berkley} at $5.7\,GHz$ the one located at a
higher bias is close to the harmonic approximation for a resonance at this
frequency.

\subparagraph{ Thrailkill et al. (2012)}

\begin{figure}[t]
\begin{center}
\scalebox{0.35}{\centering \includegraphics{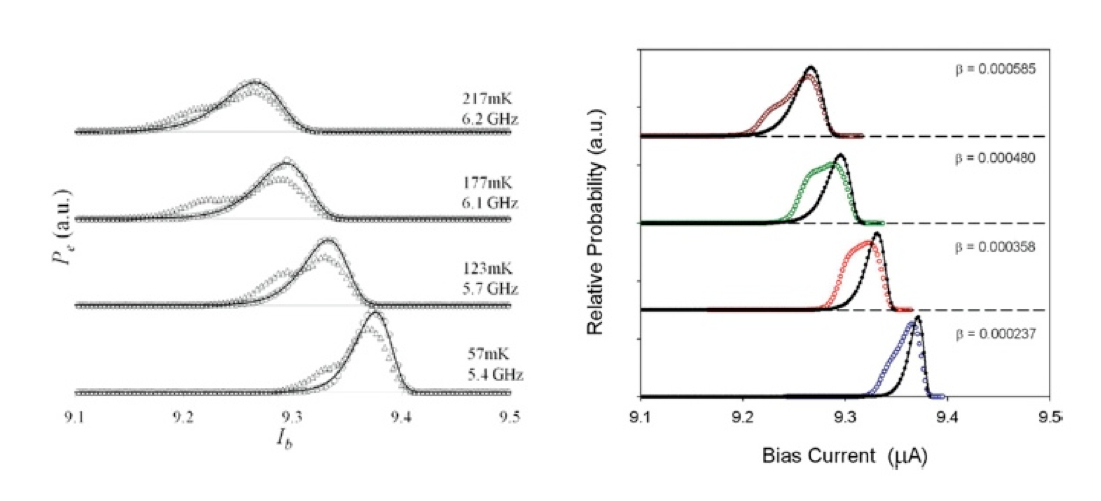}}
\end{center}
\vspace{-0.5 cm}
\caption{Left: Experimental
results for SCD peaks with (circles) and without (solid) applied microwaves.
Right: RCSJ simulation with parameters corresponding to the experiment.}
\label{Fig15}
\end{figure}

As an example, we consider an experiment by Thrailkill et al. \cite%
{Thrailkill}. In this experiment, the bias current was swept in the presence
of microwaves of specified frequency. The parameters of the junction were: $%
I_{C}=9.485\mu A$ and $C=4.7pF$; the Josephson plasma frequency was thus $%
f_{J0}=12.46GHz$. For the simulation \cite{Blackburn1}, the Lorentzian
response to the applied microwaves was specified by $a=0.09$ and $b=0.003$.
The values of the normalized bias values at resonance ($\eta _{res}$) for
each of the experimental microwave frequencies were $0.972,\,\;0.976,\,%
\,0.980,\;0.984$ respectively. As shown in figure \ref{Fig15}, the RCSJ
simulation replicates the experimental observations even in a situation
where the microwave-off peak (thermal activation) is quite close to the
microwave-on peak (ac activation).

\subparagraph{Wallraff et al. (2003)}

\begin{figure}[t]
\begin{center}
\scalebox{0.20}{\centering \includegraphics{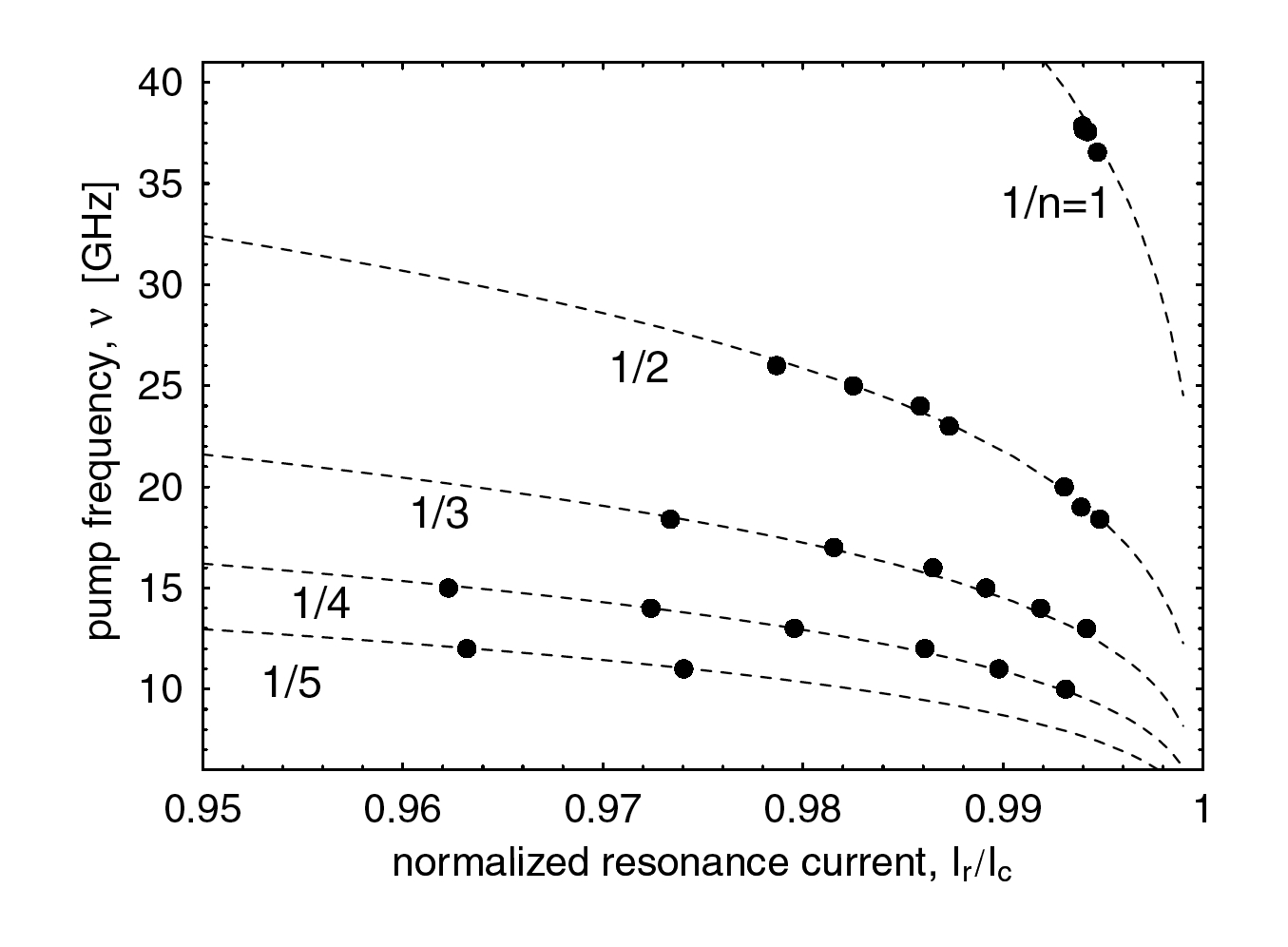}}
\end{center}
\vspace{-0.5 cm}
\caption{Applied microwave frequency $%
\protect\nu $ versus normalized resonant bias bias current $I_{r}/I_{C}$.}
\label{WallraffFig3}
\end{figure}

Wallraff et al. \cite{Wallraff2} revisited the swept bias type of experiment
pioneered in the works, almost two decades earlier, discussed in the
section: Martinis et al. (1985). As in those earlier experiments the
procedure was simple: microwaves at a given frequency were applied and
repeated bias scans were carried out to accumulate data for a switching
current distribution. The results are shown in Fig.\ref{WallraffFig3}. 
Each dot is an experimental data point. Twenty frequencies in the range $%
10.113$ - $37.894\;GHz$ were employed, \ Most were single peaks for a given
frequency, but four of these ($13.111,14.095,15.126,18.499$) yielded twin
SCD peaks separated by slightly different bias values. These experiments
showed clearly that the SCD peaks were located at bias currents for which
the frequency of the applied microwave signal was matched either to the
fundamental or to submultiple values of the expected resonance frequency
determined by that bias current according to the expression $\nu =\left(
1/n\right) \nu _{P}\left[ 1-\left( I_{r}/I_{c}\right) ^{2}\right] ^{1/4}$, $%
n=1,2,3..,$ where $I_{r}$ is the bias current at the SCD peak and $\nu
_{p}=116\;GHz$.

The frequency spacing of the quantum levels was supposed to be tuned
according to the dependence of the plasma frequency upon the external dc
bias current. It was claimed that single photon and multiphoton transitions
between junction energy levels had been observed.

However, Gr\o nbech-Jensen et al. (next section) reconsidered the
conclusions to be drawn from this experiment.

\subparagraph{Gr\o nbech-Jensen et al (2004)}

\begin{figure}[t]
\begin{center}
\scalebox{0.20}{\centering \includegraphics{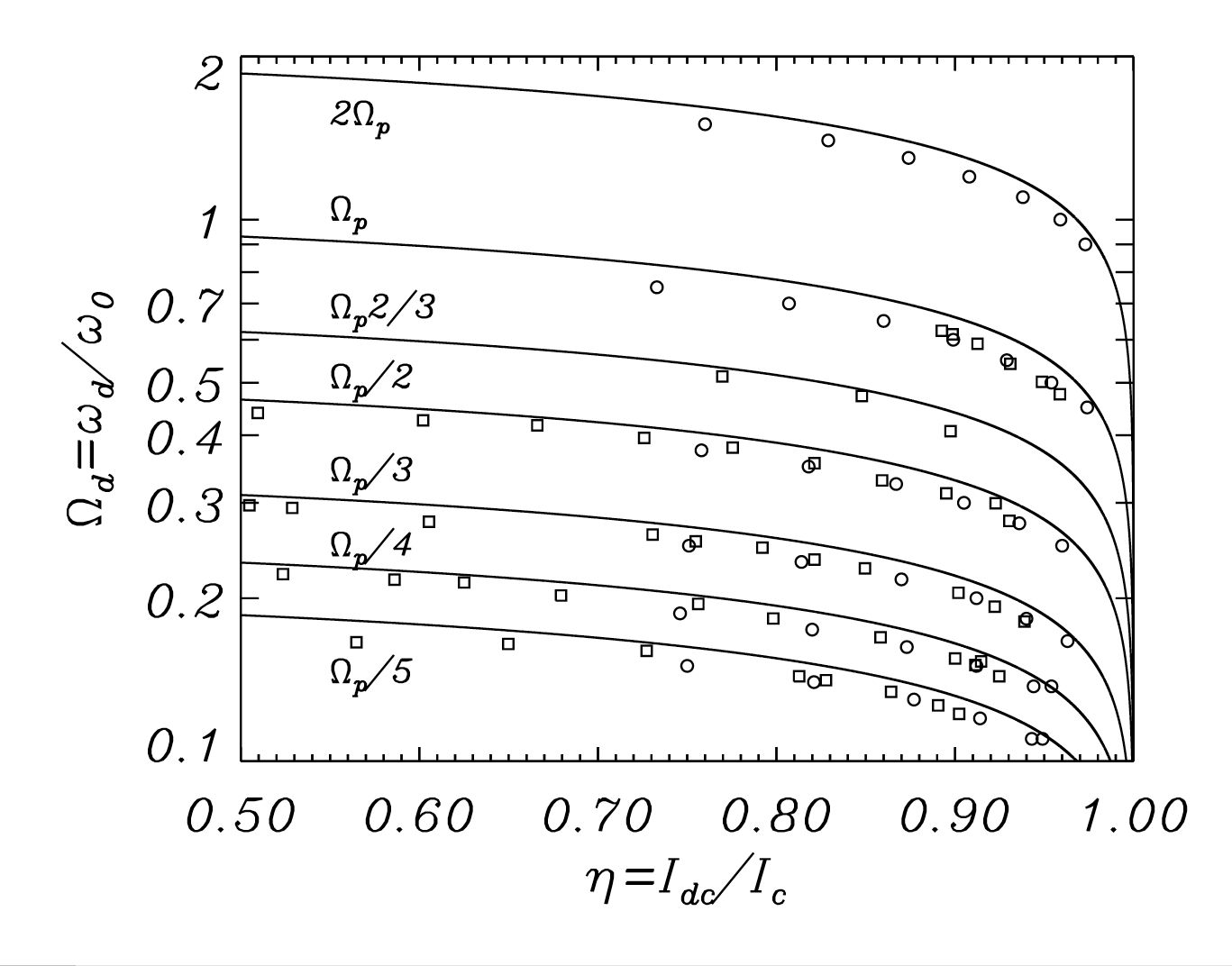}}
\end{center}
\vspace{-0.5 cm}
\caption{Positions of peaks in switching current
distributions as functions of subharmonic and harmonic pumping. Circles
represent numerical results and squares experimental data.}
\label{nielsmicrowaveharmonic}
\end{figure}

Shortly after the publication of the results of Wallraff et al. \cite%
{Wallraff2}, the collaboration of Gr\o nbech-Jensen et al. \cite{NGJ1}
reported an extensive investigation of the response of Josephson junctions
under the influence of external microwave signals in the range $350mK<T<1.5K$
. This work showed that all the excitations in the \textquotedblleft
escape\textquotedblright\ measurement spectrum could be explained, within
the RCSJ model, essentially by Eq.\ref{phase_eq} including a noise current
term; the analysis also showed striking agreement with analytical
predictions based on the interaction of the external signal with the bias
current-dependent Josephson plasma resonance. These measurements confirmed
the correctness of the data presented by Wallraff et al. \cite{Wallraff2}.
In particular, Fig. \ref{nielsmicrowaveharmonic} (Fig.3 from \cite{NGJ1})
shows the underlying harmonic features of the SCD phenomenon. 
Because numerical
simulations were in excellent agreement with the results from \cite%
{Wallraff2}, this clearly showed that the excitation spectrum belonged to
the realm of classical Josephson phenomenology. It was also shown that the
contribution of the anharmonicity of the potential well to the spectrum was
relevant \cite{NGJ2} and an excellent agreement between data and anharmonic
plasma resonance approximation was reported, as shown in Fig 4 of \cite{NGJ1}%
. Most importantly, it was found that no substantial differences exist in
the response of the junctions from $1.5K$ down to $360mK$. The latter value
was slightly above the crossover temperature of the junctions (estimated to
be $320mK$). When comparing the measurements of Refs. \cite{NGJ1, NGJ2} on
the ac response of the junctions with the similar ones collected at
temperatures safely below the crossover \cite{Wallraff2} one realizes the
complete similarity of the data and it is hard to conjecture traces of
temperature-induced discontinuities or abrupt gradients in the response of
the junctions for temperatures above, close, and below the crossover value:
this is in reality a further argument in favor of the conclusion already
expressed, that there is not much room to speculate on the existence of a
crossover temperature and a consequent transition to an MQT regime.

In the conclusion of \cite{NGJ1}, the authors state: \textquotedblleft The
experiments reported in \cite{Wallraff2} have produced ac-induced peaks in
the observed switching distributions, and the relevant peaks are located
alongside the expected classical plasma resonance curve, as we have also
found here. An important observation is that the microwave radiation
frequency necessary for populating an excited quantum level ($\hbar \omega
_{d}$) in a quantum oscillator coincides with the classical resonance
frequency of the corresponding classical oscillator. It is evident then that
multipeaked effects are not a unique signature of quantum behavior in the
ac-driven Josephson junction.\textquotedblright\ \ The last sentence is
significant in that the RCSJ was confirmed as a complete discriptor of the
experiment.

\subparagraph{Silvestrini et al.(1997)}

\begin{figure}[t]
\begin{center}
\scalebox{0.25}{\centering \includegraphics{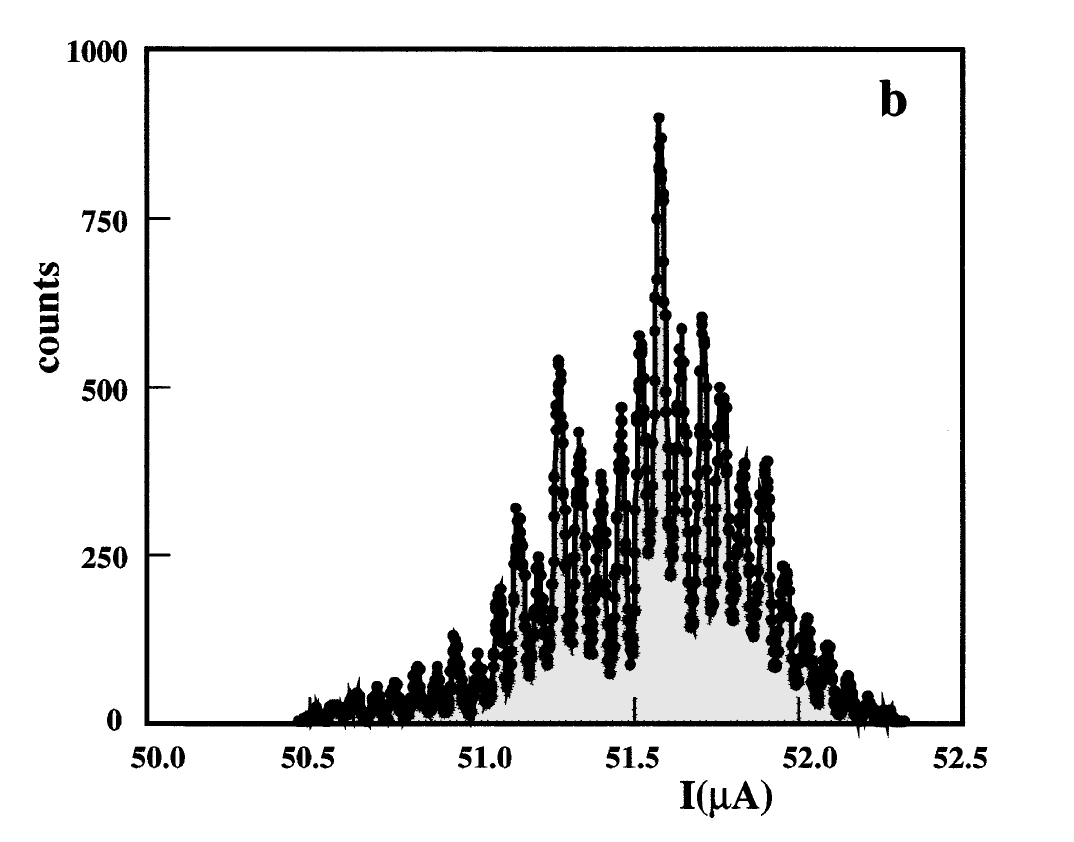}}
\end{center}
\vspace{-0.5 cm}
\caption{Figure 2 b from \protect\cite{Silve97} showing
experimental switching current distribution at high sweep rate.}
\label{silvestriniFig2}
\end{figure}

An interesting experiment claiming the observation of quantized energy
levels of the washboard potential, in the absence of applied microwave
radiation, was reported in 1997 by P. Silvestrini et al..\cite{Silve97} The
idea of this experiment was to evidence quantum levels through a very fast
ramp of the current biasing the junctions. The physics at the basis of the
experiment was that it should be easier to observe possible quantized levels
of the potential wells when a fast ramping out of these generates non
stationary conditions which do not give the possibility to the junctions to
thermalize. In principle, the experiments could be performed even above the
crossover temperature and in fact the measurements were performed in the
range $1.5K-4.2K$. This idea had a significant impact on the community
because the results, see Fig.\ref{silvestriniFig2}, 
appeared to represent an independent proof of the existence of the quantized
levels previously claimed by the microwave spectroscopy experiments.

\begin{figure}[t]
\begin{center}
\scalebox{0.30}{\centering \includegraphics{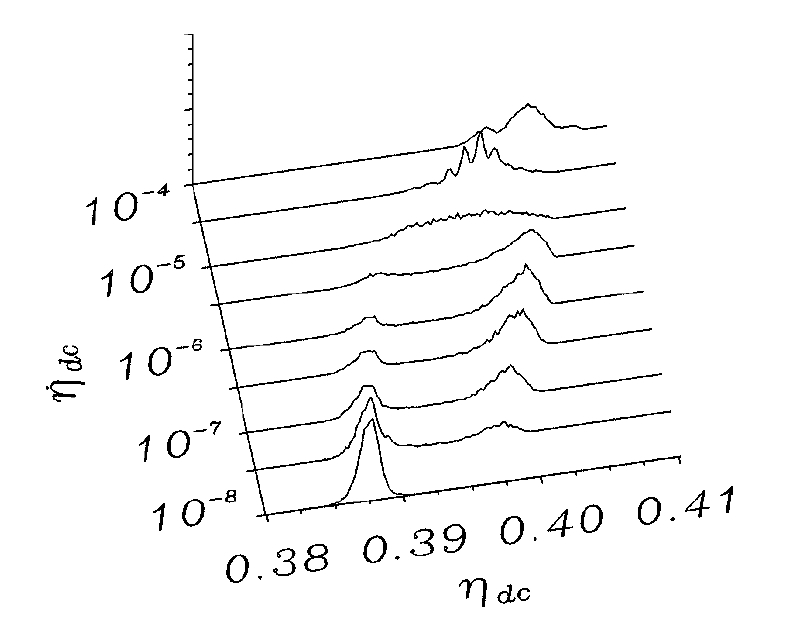}}
\end{center}
\vspace{-0.5 cm}
\caption{Simulated
switching distributions at different sweep rates ($\dot{\protect\eta}%
_{dc}=10^{-8}\rightarrow 10^{-4}$).}
\label{nielsmatteoannularfig7}
\end{figure}

We know now that the microwave spectroscopy experiments of the 80's can be
well accounted for by RCSJ model physics, and so a claim of unambiguous
evidence of quantized levels should be based exclusively on the results of
the Silvestrini 1997 experiment. Such a claim would also be questionable
since we have observed several times, both in small area and large area
junctions RCSJ simulations, that a very fast bias current ramp can generate,
in thermal escape histograms, effects that look like the \textquotedblleft
levels\textquotedblright\ reported by Silvestrini et al. \cite{Silve97}. In
Fig.\ref{nielsmatteoannularfig7} from \cite{NGJ6} one can see a typical
example of the dependence of escape histograms upon the sweep rate: a multi
levels-like histogram can be seen in this specific case for $10^{-5}<\frac{d%
}{dt}\eta _{dc}<10^{-4}$ 
For high sweep rates
one can obtain features very similar to those of Fig.\ref{SilvestriniFig2};
moreover, we have observed that the "levels" exhibited in the RCSJ model at
fast sweeps also disappear with increasing losses or temperature. Thus, even
in the case of the experiments of Silvestrini et al.\cite{Silve97} the
observed effects cannot be uniquely attributed to quantum modelling. The
sweep rate is a relevant parameter as far as escape from potential wells is
concerned and its relation with the statistical output of the
experiments/simulations must be evaluated very carefully.

\subsection{Annular Junctions}

\begin{figure}[t]
\begin{center}
\scalebox{0.25}{\centering \includegraphics{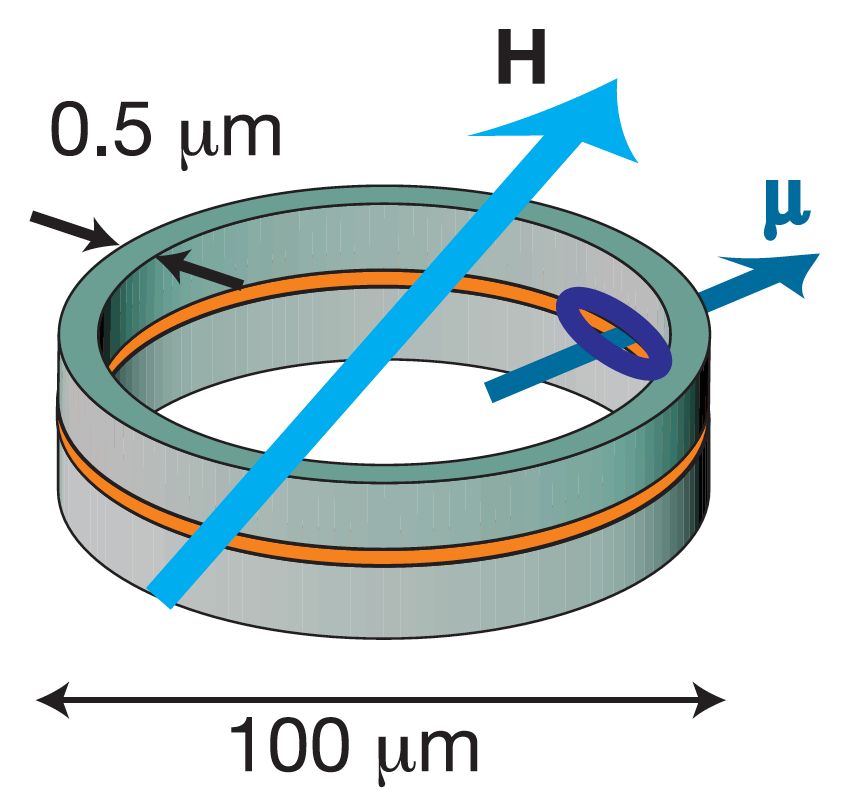}}
\end{center}
\vspace{-0.5 cm}
\caption{Schematic diagram of an annular Josephson
junction. Superconducting electrodes (not shown) connect to the upper and
lower portions of the junction. \textbf{H} is an externally applied magnetic
field and $\mathbf{\protect\mu }$ indicates the magnetic moment of a single
trapped fluxon.}
\label{annularjunction}
\end{figure}

\begin{figure}[t]
\begin{center}
\scalebox{0.4}{\centering \includegraphics{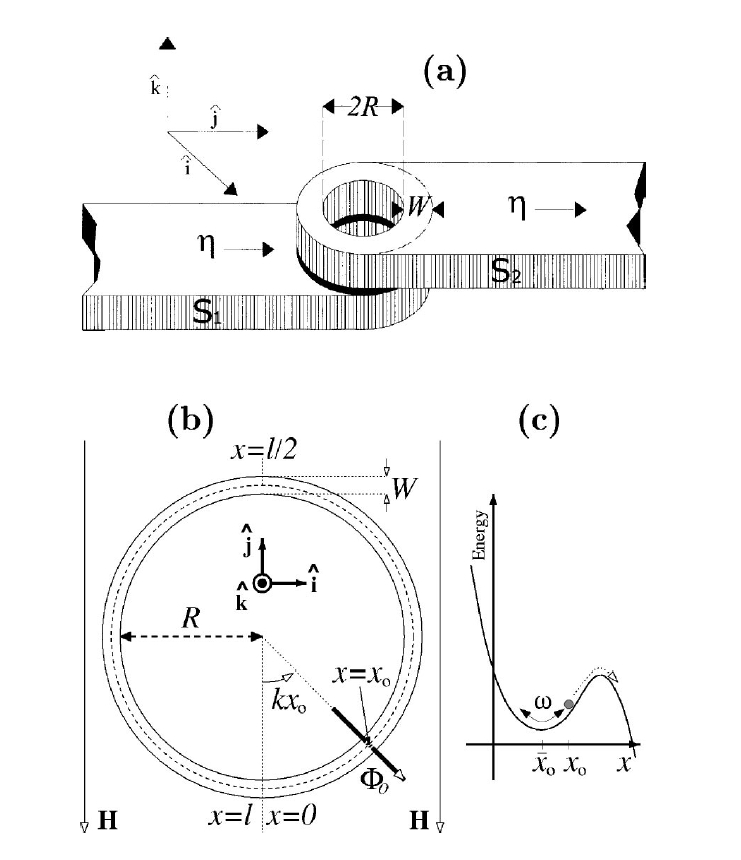}}
\end{center}
\vspace{-0.5 cm}
\caption{Fig. 1 from \protect\cite{NGJ6}
showing a complete annular junction.(a) and the potential energy as a
function of the vortex coordinate.}
\label{completeannular}
\end{figure}

The preceding sections have considered phase dynamics in small Josephson
junctions, namely junctions whose spatial dimensions are small with respect
to the Josephson penetration depth $\lambda _{j}=\sqrt{\frac{\Phi _{0}}{2\pi
\mu _{0}dJ_{c}}}$ (here $\mu _{0}=4\pi 10^{-7}H/m$ , $d$ the magnetic
thickness of the junction and $J_{c}$ is the Josephson maximum super current
density).\ We now turn our attention to another, closely related, system -
an annular Josephson junction- namely a junctions for which the
superconducting electrodes overlap along a circle. In this case we have just
one spatial degree of freedom for the phase if the length of the circle is
larger than $\lambda _{j}$ while the overlapping region is smaller than $%
\lambda _{j}$; it is known that this degree of freedom, even just in a
one-dimensional configuration, gives rise to the presence of Josephson
flux-quanta (fluxons) along the spatially extended dimension, i.e., along
the circular path \cite{Davidson1986, Ustinov, Martucciello}; an annular
junction is depicted in Fig.\ref{annularjunction} \cite{Wallraff}.
where we can see ring-shaped upper
and lower superconductors separated by an oxide barrier. The complete device
can be visualized as the overlap junction shown in Fig.\ref{completeannular}%
with a hole drilled
through the overlapping fingers as indicated in Fig. \ref{completeannular}.
Bias current is sent in to and out from the junction via superconducting
films.

The experiments of Wallraff et al. \cite{Wallraff} observed the motion of a
single trapped fluxon in the plane of the annular junction - depicted as a
vector labelled $\mathbf{\mu }$ in the figure. The technique for trapping a
single fluxon in the ring is discribed in \cite{Ustinov}: \textquotedblleft
trapping of a magnetic flux in the junction ring was made while cooling the
sample below the critical temperature $T_{C}=9.2$ $K$ of
niobium\textquotedblright . As noted in Martucciello et al. \cite%
{Martucciello}, about $10$ or $20$ attempts could be necessary before a
single fluxon was trapped this way.

A fixed external magnetic field $\mathbf{H}$ is applied in the plane of the
junction; this sets the preferred minimum energy orientation of the fluxon,
just as gravity does for a free hanging pendulum. A steady bias current
passes vertically through the annular junction as a ring shaped sheet. \
Since the fluxon axis is always radially directed, the Lorentz force on the
fluxon is always tangent to the ring, and is constant. This is analogous to
a steady torque on a pendulum. Whereas a pendulum has a critical torque ($%
mg\ell $) beyond which rotational motion occurs, the annular junction will
have a critical bias current above which the fluxon will circulate around
the ring. \ When that happens, a voltage will appear across the junction.

The dynamics of this system are described in \cite{NGJ6}.%
\begin{equation}
\varphi _{tt}-\varphi _{xx}+\sin \varphi =\eta _{dc}+\eta _{ac}\sin \omega
_{d}t+\Gamma k\sin kx-\alpha \varphi _{t}+n(x,t)  \label{sineGordon}
\end{equation}%
where $\varphi (x,t)$ is the order parameter phase difference across the
junction and $x$ is the fluxon coordinate measured around the circumference
of the ring. This is a sine-Gordon equation with dissipation $\alpha $, a
forcing term due to an applied magnetic field $\Gamma $ and, finally,
thermal noise $n(x,t)$.

\subparagraph{Experiment with Microwaves OFF}

\begin{figure}[t]
\begin{center}
\scalebox{0.2}{\centering \includegraphics{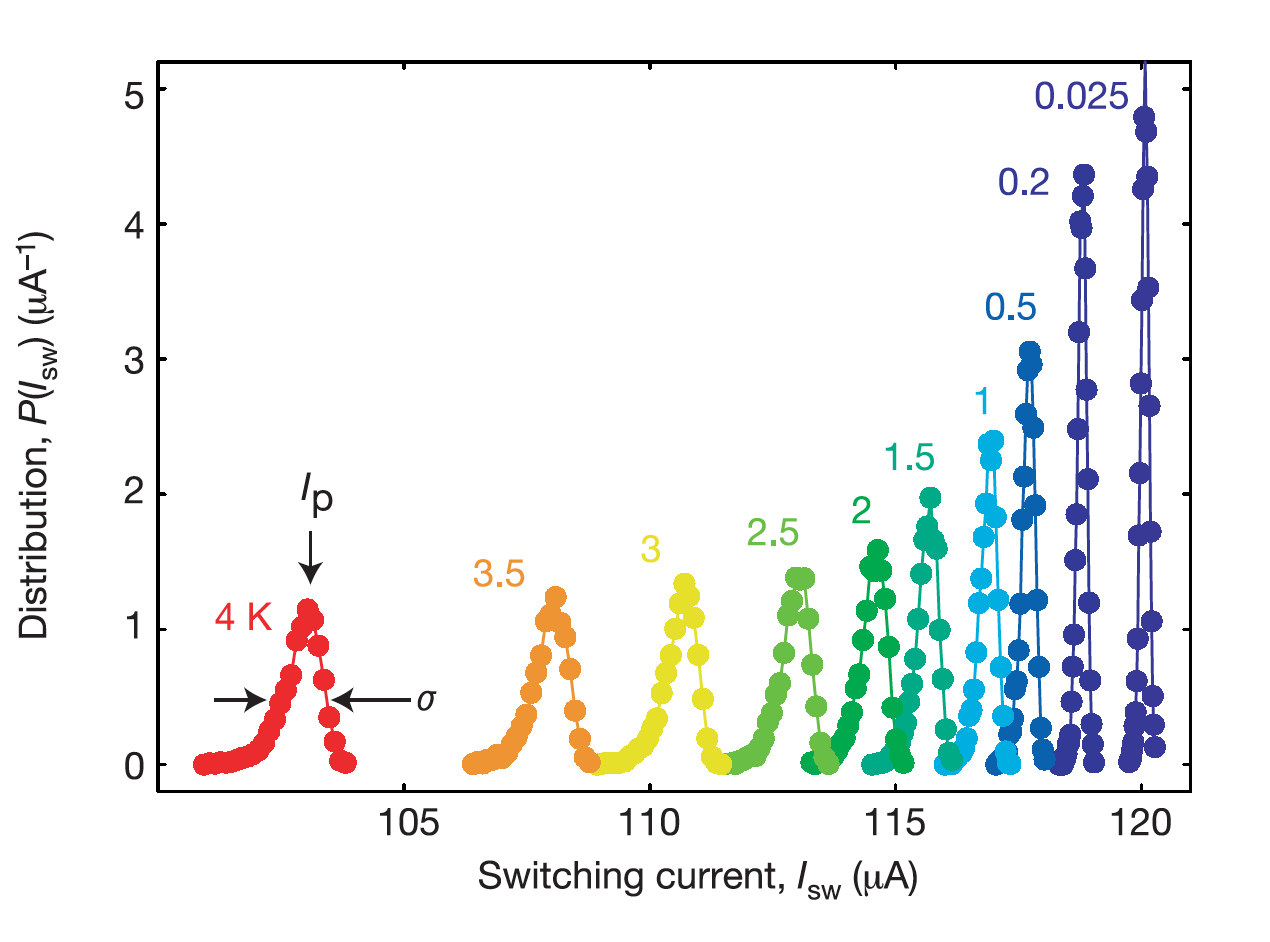}}
\end{center}
\vspace{-0.5 cm}
\caption{Probability distributions for escape to a running fluxon state in an
annular Josephson junction \protect\cite{Wallraff}.}
\label{Wallraff2a}
\end{figure}

As with a \ simple Josephson junction, the point at which the switch from
static (zero voltage) to dynamic (nonzero voltage) occurs depends on the
presence of any additional noise and, in these experiments, this is thermal
noise. Noise causes the switching to happen just before the critical bias is
reached, but with statistical likelihoods. As with the simple Josephson
junction, repeated bias sweeps produce distributions in the switching moment
(see Fig.\ref{Fig3}). The temperature dependence of this escape process is
evident from Fig.\ref{Wallraff2a}. 
The peaks in the swept bias
escape distributions clearly have the same attributes as the corresponding
data for a small area Josephson junction (Figs.\ref{Fig3} and \ref{Fig4}).

\begin{figure}[t]
\begin{center}
\scalebox{0.2}{\centering \includegraphics{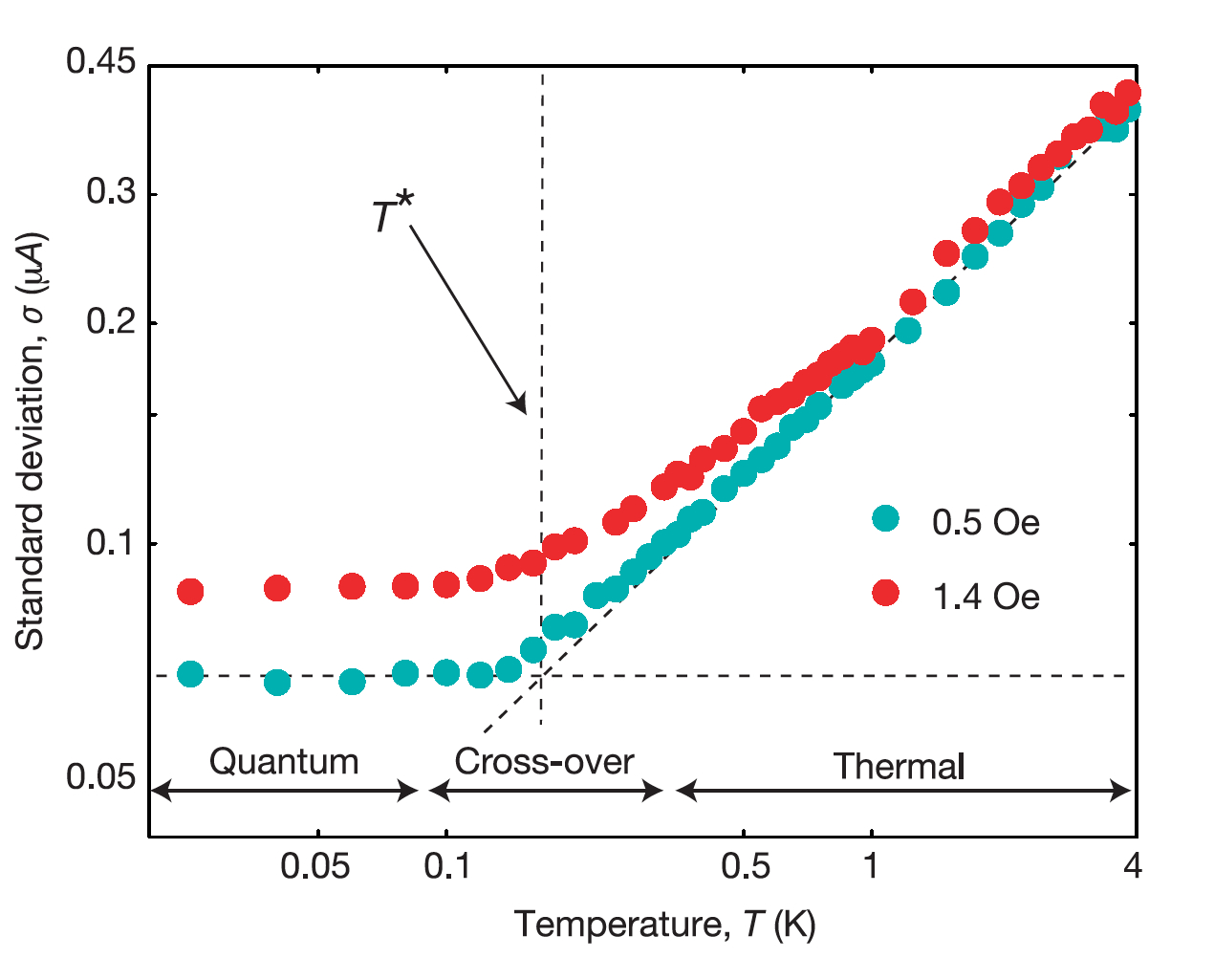}}
\end{center}
\vspace{-0.5 cm}
\caption{Figure 2 from \protect\cite%
{Wallraff} showing the SCD escape peak widths as a function of temperature.}
\label{WallraffWidth}
\end{figure}

As is commonly done, peak widths were chosen for close scrutiny and
presented on a log-log plot (\cite{Wallraff}, Fig.\ref{WallraffWidth}).
There does appear to be a suggestion of saturation of the widths at the
lowest lowest temperatures, but as discussed in III A, the log-log treatment
gives a visual appearance of flattening that may be deceiving. \ 

\begin{figure}[t]
\begin{center}
\scalebox{0.3}{\centering \includegraphics{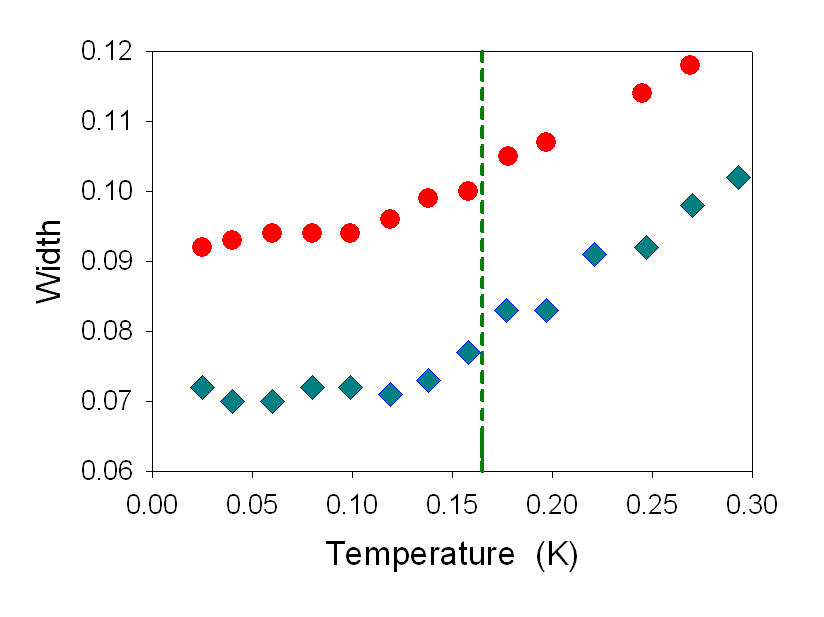}}
\end{center}
\vspace{-0.5 cm}
\caption{Data from \protect\cite{Wallraff}, Fig 2b plotted with
linear scales. The dashed line indicates the postulated crossover
temperature.}
\label{wallraffwidth}
\end{figure}

For example, the same data in this figure can be replotted with conventional
linear scales, as is done in Fig.\ref{wallraffwidth}.
Now the trends appear
gradual, without any dramatic flattening below the purported crossover
temperature. However, peak widths are quite small at the lowest temperatures
and their precision is much less than peak position data.

\begin{figure}[t]
\begin{center}
\scalebox{0.3}{\centering \includegraphics{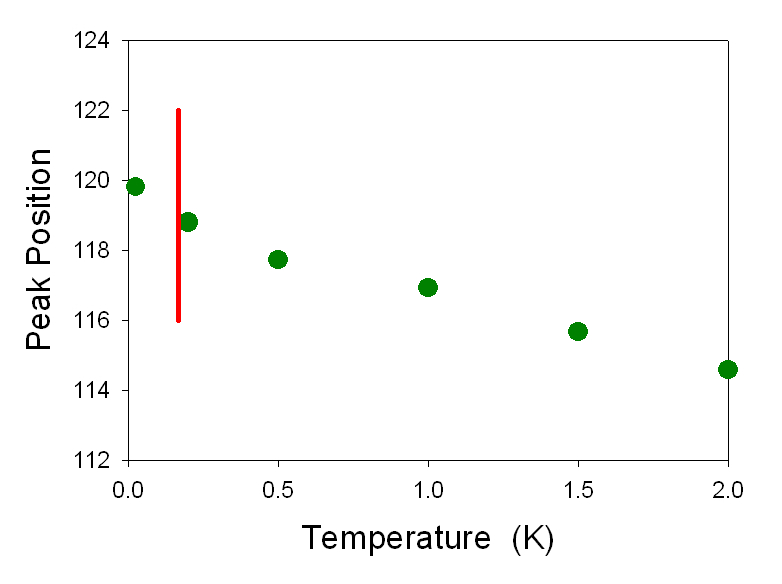}}
\end{center}
\vspace{-0.5 cm}
\caption{Peak positions from 
\protect\cite{Wallraff} as functions of the bath temperature. \ The vertical
line indicates the reported crossover temperature.\ Note the absense of peak
`freezing' at the lowest temperatures.}
\label{wallraffposition}
\end{figure}

More revealing is the behavior of peak positions. As with the data of Voss
and Webb \cite{VossWebb} and Yu et al. \cite{Yu}, escape peaks reported by
Wallraff et al. (\cite{Wallraff} Fig. \ref{Wallraff2a}) are seen to advance
steadily as the temperature is lowered, and this continues well below the
anticipated crossover temperature. It is clear in Fig.\ref{wallraffposition}
that this trend does not exhibit any sign of becoming
temperature-independent as would be expected if there were a transition to
MQT where the escape rate itself would be temperature independent. (It would
be helpful if many more data points were recorded to fill in this plot below 
$500$ mK) 
This alone calls into
question any claim that a condition of temperature-independence has been
reached at the lowest temperatures of the experiment.

The trend in these experimental widths is unconvincing, and the trend in the
experimental positions decidedly disagrees with quantum expectations.

\subparagraph{\ Experiment with Microwaves ON}

The focus of experiments on annular junctions has been the effects of
applied microwaves. From a classical phenomenological perspective, it would
be anticipated that resonant activation would occur when the microwave
frequency matched the natural frequency of a fluxon at a given bias
condition.

The experimental results from \cite{Wallraff} are shown in Fig.\ref%
{wallrafffig3a}.

\begin{figure}[t]
\begin{center}
\scalebox{0.25}{\centering \includegraphics{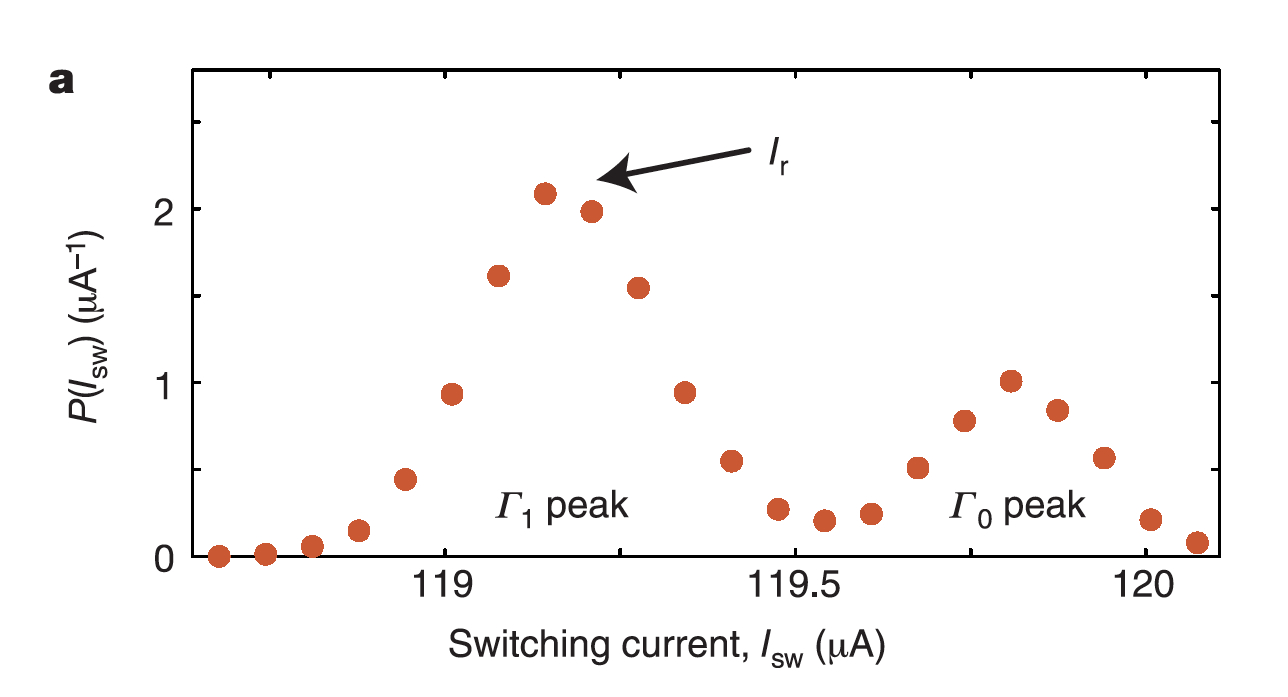}}
\end{center}
\vspace{-0.5 cm}
\caption{SCD escape peaks with $11$ GHz
applied microwaves at $T=25$ mK.(Fig.3a from \protect\cite{Wallraff}).}
\label{wallrafffig3a}
\end{figure}
According to
Wallraff et al.: \textquotedblleft The peak at higher bias current is due to
the tunnelling of the vortex from the ground state, whereas the peak at
lower bias current corresponds to the tunnelling out of the first excited
state\textquotedblright . In other words, the $\Gamma _{0}$ peak is present
with or without microwaves whereas the $\Gamma _{1}$ peak is observed
additionally when $11$ GHz microwaves are present. The peaks are at
switching currents of $\ \Gamma _{0}=119.79\mu $A and $\Gamma _{1}=119.09\mu 
$A; $\ \Gamma _{0}$ is the same as the $25$ mK peak in Fig.\ref{Wallraff2a}.
\ Note the similarity of this two-peak plot to the data of Thrailkill et al. 
\cite{Thrailkill} shown in Fig.\ref{Fig15} where one peak is microwave
induced and the other is simple thermal activation.

\begin{figure}[t]
\begin{center}
\scalebox{0.5}{\centering \includegraphics{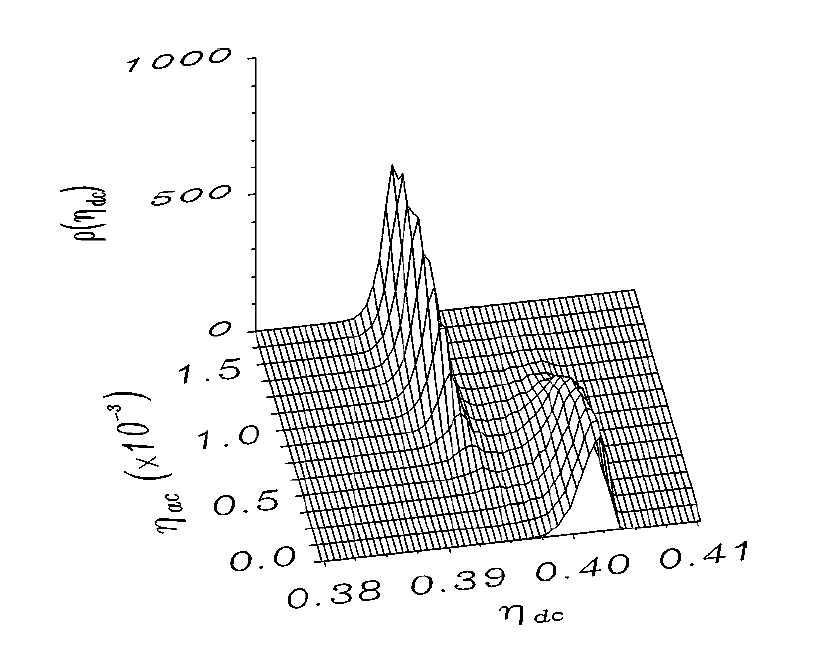}}
\end{center}
\vspace{-0.5 cm}
\caption{Simulation results for an annular Josephson junction \protect\cite%
{NGJ6}. \ For each possible choice of ac drive amplitude, the vertical axis
indicates the relative probability that the junction will switch to a finite
voltage state (circulating fluxon) as a function of dc bias.}
\label{NielsAnnularFig2}
\end{figure}

Numerical simulations based on Eq.\ref{sineGordon} yield results such as
depicted in Fig.\ref{NielsAnnularFig2}. 
With no excitation $\eta _{ac}=0$, there is a single escape peak. It is also
evident that there is a range of excitation amplitude $\eta _{ac}$ for which
there are two SCD peaks, the upper one that occurs even without excitation
and a second peak at a lower $\eta _{dc}$ dc current. \ For $\eta
_{ac}\simeq 0.8$ this double peak structure approximately matches the
experimental result shown in Fig.\ref{wallrafffig3a} and confirms the
adequacy of the classical RCSJ model in capturing the system behavior.

\section{\protect\large Artificial Atom Analogies}

\subsection{Rabi Oscillations}

At the end of the 90's a general belief existed that robust evidences of new
macroscopic quantum coherence effects in Josephson systems had been achieved
(indeed unambiguous evidences were already claimed at the end of the 80's 
\cite{Leggett87}). Along with the fact that in the mid-90's two theoretical
algorithms showed the powerful possibilities offered by systems operating in
a controllable quantum superposition state \cite{Shor, Grover} the above
evidences generated a noticeable squeezing of intellectual resources toward
experiments which could demonstrate possibility of manipulation of the
Josephson quantum states. In 2002 John M. Martinis and his group in Boulder 
\cite{MART2}, relying on the fact that a Josephson junction could behave as
system with quantized energy levels, presented results in which the physics
predicted for a two levels quantum systems could be expected from a
Josephson junction and its washboard potential. The physics of the
fundamental process was dating back to the arguments presented by Isaac Rabi
in 1937 \cite{Rabi37} who considered a multilevel atomic system. Rabi
oscillations have been also considered within the context of quantum optics 
\cite{Loudon}.

The application of external microwaves of adequate frequency to the atom
(the atom being the Josephson junction) for a given time would generate a
flipping between energy levels which, in turn, could give rise to
oscillations in the population of the energy levels, known as Rabi
oscillations. The status of the system (i.e., the level population) was
investigated by a technique employing a sequence of pump pulses and probe
pulses. Both pump pulse and probe pulse consisted of a small amplitude
microwave signals applied to the junctions for given times: the pump pulse
was responsible for the flipping between levels while the probe pulse was
responsible for testing the final population in the states by potential
escape techniques.

The evidence of Rabi oscillations\cite{MART2}, namely the oscillations in
the population of the first excited level as a function of the applied
microwave power was striking and therefore this experiment was considered a
conclusive evidence of the possibility that Josephson junctions could behave
as elementary bits of quantum information, or qubits. Indeed, from that time
on a new terminology took over when referring to Josephson circuits: in
particular, circuits relying on single tunnel junctions for quantum
operation were baptized \textquotedblleft phase qubits\textquotedblright\
while circuits employing macroscopic quantum flux states (like rf-SQUIDSs
and dc-SQUIDs) \textquotedblleft flux qubits\textquotedblright . Beside the
fashionable names it is worth recalling these were just Josephson junctions
devices. The application of microwave pulses of suitable frequency and
duration to an atom could increase the population of an upper state. In the
Rabi effect the degree of population enhancement is a periodic function of
the amplitude of the pump pulse. The status of the system (i. e., the level
population) can be investigated by means of a probe pulse. Both pump pulse
and probe pulse consist of a small amplitude microwave signals applied for
given times. \ The experiments of Martinis et al. \cite{MART2} were
predicated on the supposition that a Josephson junction at temperatures
around $50$ $mK$ would have entered a macroscopic quantum state, and thus
could be viewed as an artificial atom. \ Therefore, phenomena like Rabi
oscillations ought to be observable.

\begin{figure}[t]
\begin{center}
\scalebox{0.35}{\centering \includegraphics{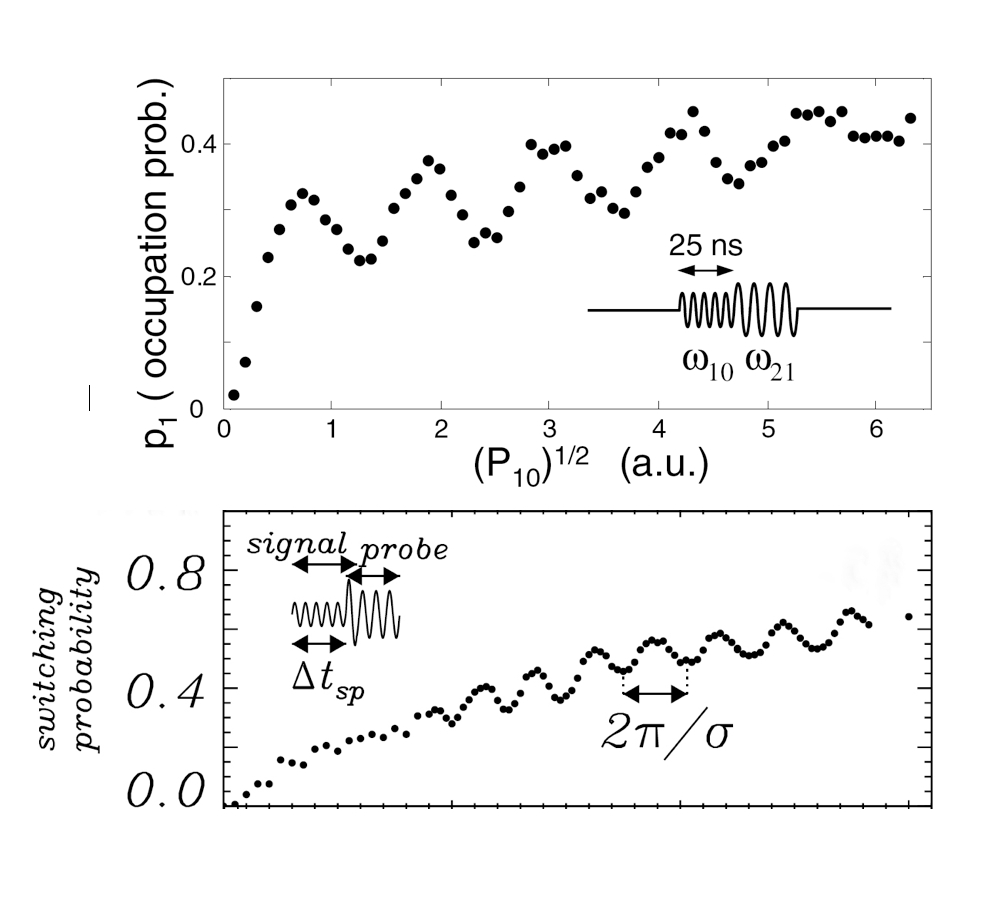}}
\end{center}
\vspace{-0.5 cm}
\caption{Rabi oscillations: upper panel, experimental results
reported in \protect\cite{MART2}. \ The probability is plotted as a function
of the microwave pulse amplitude; lower panel, simulation based on classical
junction description \protect\cite{NGJ4}.}
\label{RabiDoublePlot}
\end{figure}

The experimental evidence for Rabi oscillations presented in \cite{MART2}
(see the upper panel of Fig.\ref{RabiDoublePlot})
seemed persuasive and this
added to the consensus in the MQT community that a Josephson junction would
indeed function as a qubit at very low temperatures.

\begin{figure}[t]
\begin{center}
\scalebox{0.25}{\centering \includegraphics{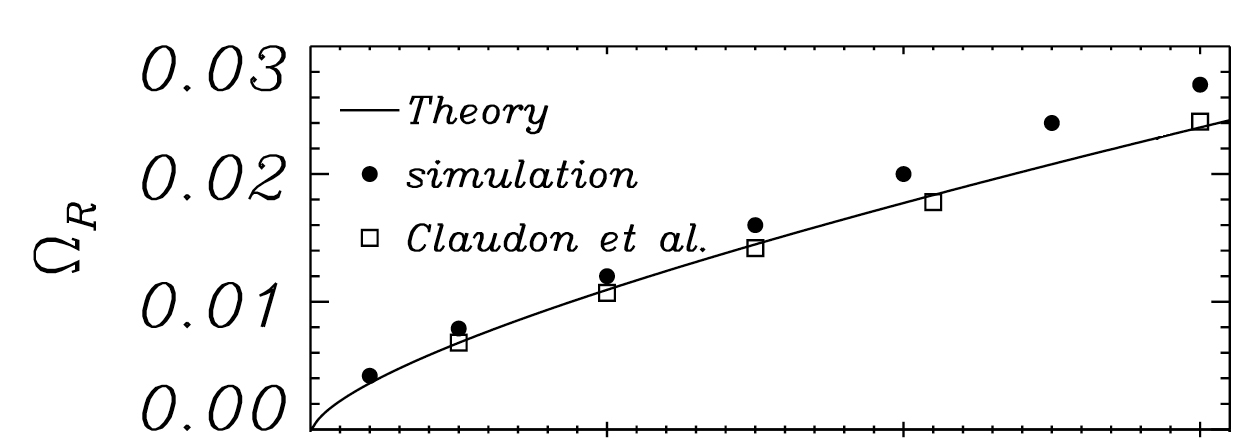}}
\end{center}
\vspace{-0.5 cm}
\caption{Modulation frequency $\Omega _{R}$
as a function of microwave amplitude $\protect\epsilon _{S}$ (horizontal
scale $\protect\epsilon _{S}:0\rightarrow 0.003$).}
\label{nielsrabi2005}
\end{figure}

During the first half of the 2000-2010 decade, however, numerical techniques
for simulating escape processes in Josephson junction through the RCSJ model
had been set up and calibrated in a number of papers \cite{Kautz, NGJ0}. The
numerical simulations had shown striking agreement with experimental reality
and analytical approaches for ac driven junctions. Thus, three years after
the appearance of Martinis and co-workers' paper on the Rabi oscillations,
papers were published \cite{NGJ3,NGJ4} in which it was shown that the
features reported in the Martinis Rabi paper could be well interpreted by
the nonlinear phenomenology of Josephson junctions. The analytical
approximations were in striking agreement with numerical simulations and the
experimental results \cite{MART2, Claud1} as revealed in the lower panel of
Fig.\ref{RabiDoublePlot}, and by the comparison between analytical ansatz,
numerical simulation and experiments shown in \ Fig.\ref{nielsrabi2005}. 
The basis for the
\textquotedblleft classical\textquotedblright\ explanation of Rabi-like
oscillations in Josephson junctions was a phase-locking ansatz (between
junction internal oscillations and microwave signal) and a perturbation
analysis of the phase-locked states introducing a transient modulation whose
frequency could capture all the observed experimental features.

\subsection{Ramsey fringes and Spin echoes}

\begin{figure}[t]
\begin{center}
\scalebox{0.4}{\centering \includegraphics{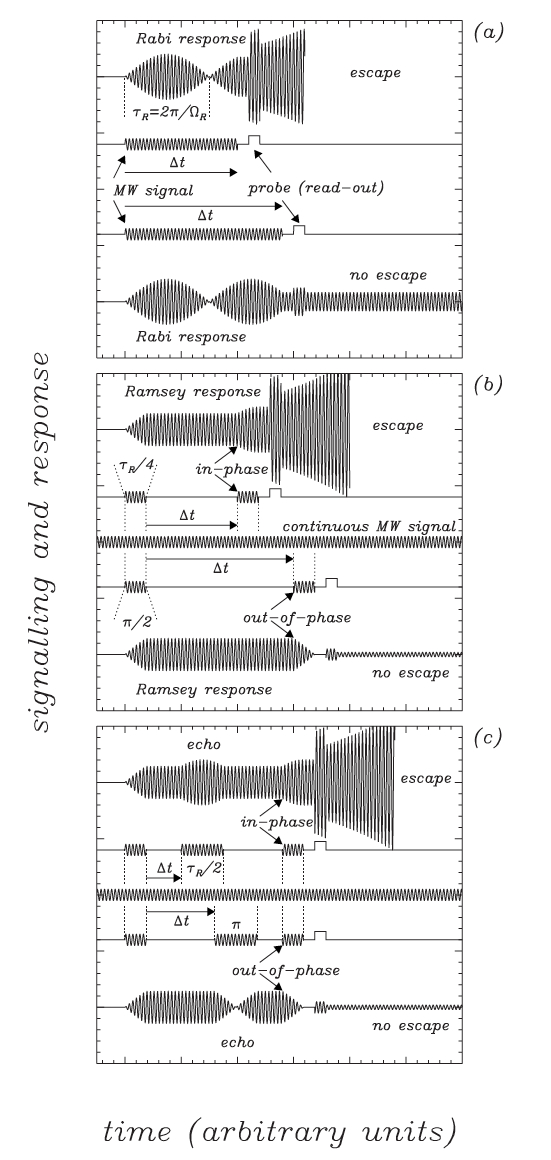}}
\end{center}
\vspace{-0.5 cm}
\caption{Figure 4 from \protect\cite{JEFF3}.}
\label{JEFF3FIG4}
\end{figure}

Following the paper by Martinis et al. in 2002 other publications appeared 
\cite{EstDev, PLOURDE05} in which the experimental \textquotedblleft
microwave pulses\textquotedblright\ protocol was extended to reproduce other
features (mostly coming from analogies with NMR research techniques) of the
Josephson artificial atom like the observation of Ramsey fringes \cite%
{Ramsey50} and spin echoes \cite{Hahn50}. However, it was soon demonstrated
that even these features could be explained as consequences of the RCSJ
model of Josephson systems. Both Ramsey fringes \cite{JEFF1} and spin echo
oscillations were reproduced for a single junction system \cite{JEFF2} and
for SQUIDs \cite{JEFF3}. Ramsey fringes and spin echo--like effects could be
activated in Josephson junction systems by appropriate combined sequences of
pumping and probe pulses. These combinations can mimic accurately what were
attributed to rotations on the Bloch sphere\cite{EstDev}: A synthetic scheme
for the generation of Rabi oscillation, Ramsey fringes and spin echo are
shown in Fig. \ref{JEFF3FIG4} (Fig.4 of \cite{JEFF3}).

\begin{figure}[t]
\begin{center}
\scalebox{0.35}{\centering \includegraphics{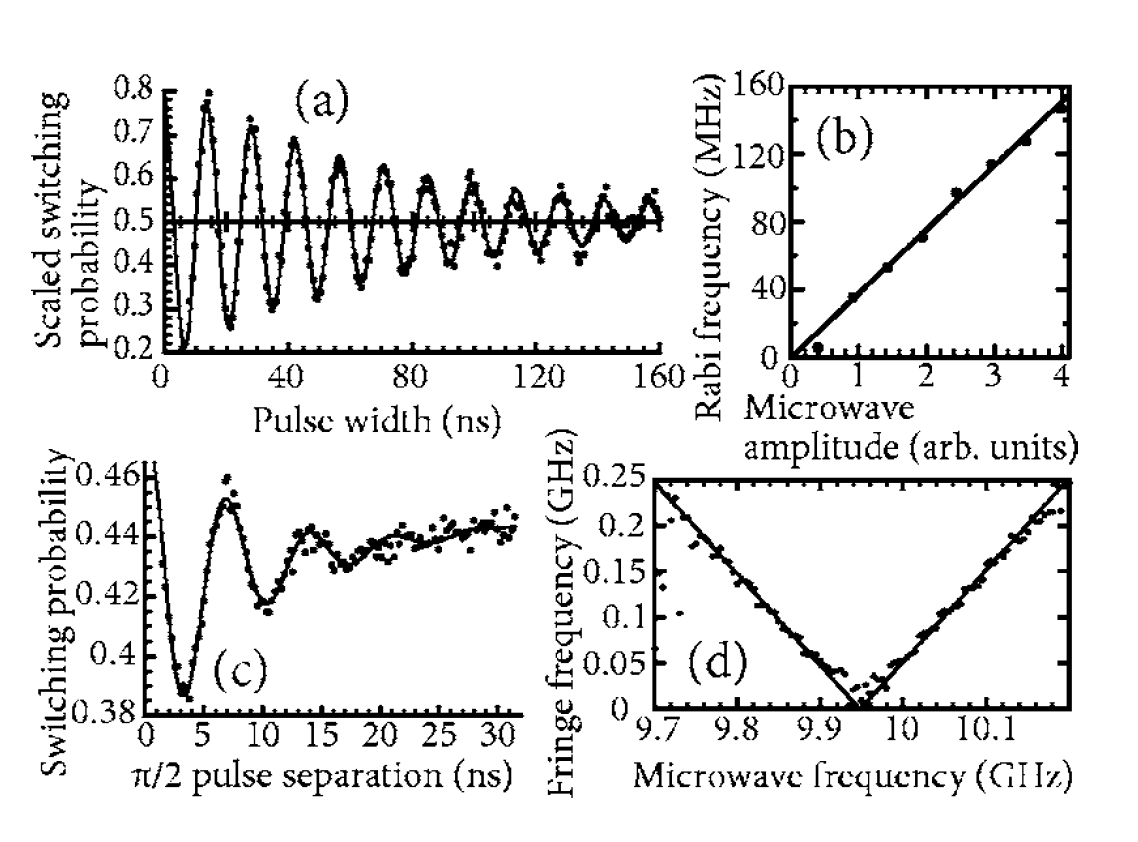}}
\end{center}
\vspace{-0.5 cm}
\caption{Figure 4 from \protect\cite{PLOURDE05}.}
\label{PLOURDEFIG4}
\end{figure}

In Fig. \ref{PLOURDEFIG4} (Fig 4 from \cite{PLOURDE05}) \ we show Rabi
oscillations and Ramsey fringes measured on a SQUID system investigated from
the \textquotedblleft quantum\textquotedblright\ point of view,
(Fig.5 from \cite{JEFF3}) 
\begin{figure}[t]
\begin{center}
\scalebox{0.35}{\centering \includegraphics{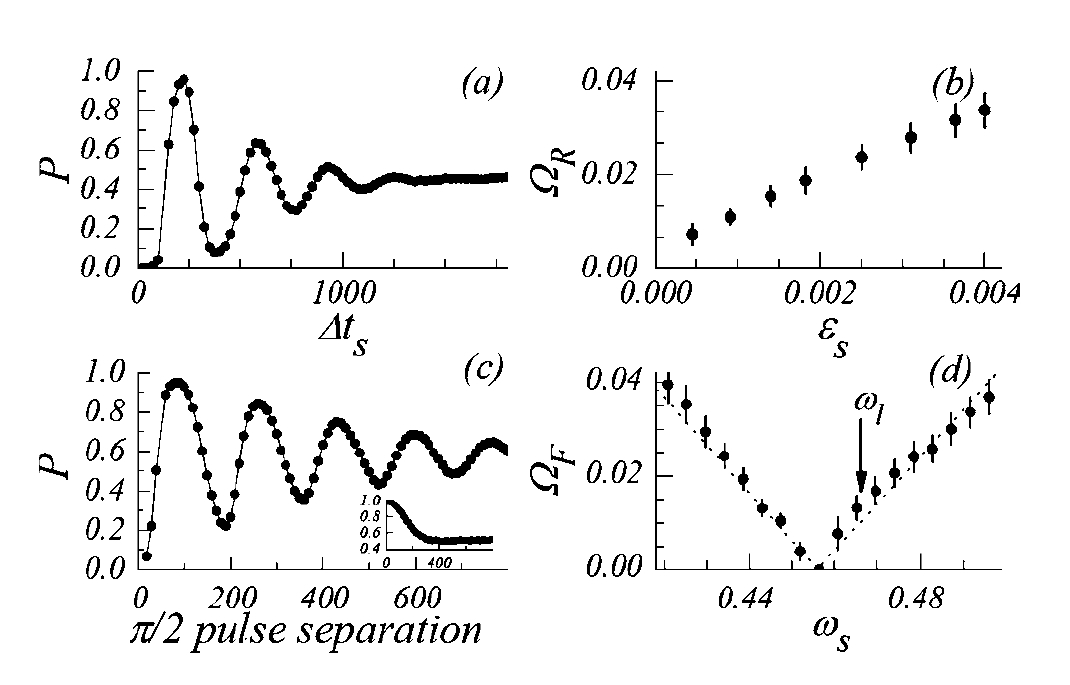}}
\end{center}
\vspace{-0.5 cm}
\caption{Figure
5 from \protect\cite{JEFF3}.}
\label{JEFF3FIG5}
\end{figure}
we show results of RCSJ
simulations of the same system. As can be seen in these cases, like in all
the others already discussed, the agreement of the RCSJ output with the
experiments is remarkable.

The experiment described in \cite{MART2} was framed by the idea that the
washboard potential for a Josephson junction would become quantized at very
low temperatures and therefore, having thus become an artificial atom, it
should exhibit phenomena such as Rabi oscillations. \ The very names
\textquotedblleft pump\textquotedblright\ and \textquotedblleft
probe\textquotedblright\ in speaking of the weak microwave bursts convey the 
\textit{expectation} of what would be occurring in the experiment - that
some of the occupants of a lower state of the artificial atom would be
pumped up to a higher state by a microwave pulse, and that the altered
occupancy of the higher level could be sensed with a second probe pulse.

Oscillations that look like Rabi oscillations may indeed spring from another
type of physics and thus do not constitute proof that anything quantum is
going on.

\section{\protect\large Coupled Junctions}

\begin{figure}[t]
\begin{center}
\scalebox{0.5}{\centering \includegraphics{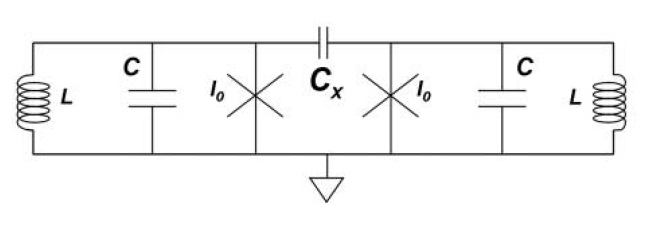}}
\end{center}
\vspace{-0.5 cm}
\caption{Diagram from \protect\cite%
{Steffen} representing the equivalent circuit of the experimental system.}
\label{Fig16}
\end{figure}

If superconducting qubits are to be exploited for quantum computing, it
needs to be possible to produce entangled states. To this end, experiments
were devised (\cite{Steffen}) in which a pair of qubits, macroscopically
coupled by a circuit element -- a capacitor -- were subjected to a sequence
of microwave bursts and pulses intended to condition and examine the state
of the qubits. Our approach to a description of this experiment is strictly
from the point of view of the RCSJ junction model. Steffen et al. \cite%
{Steffen} provided the schematic shown in Fig.\ref{Fig16} as an equivalent
circuit for their coupled experimental qubits.

\begin{figure}[t]
\begin{center}
\scalebox{0.25}{\centering \includegraphics{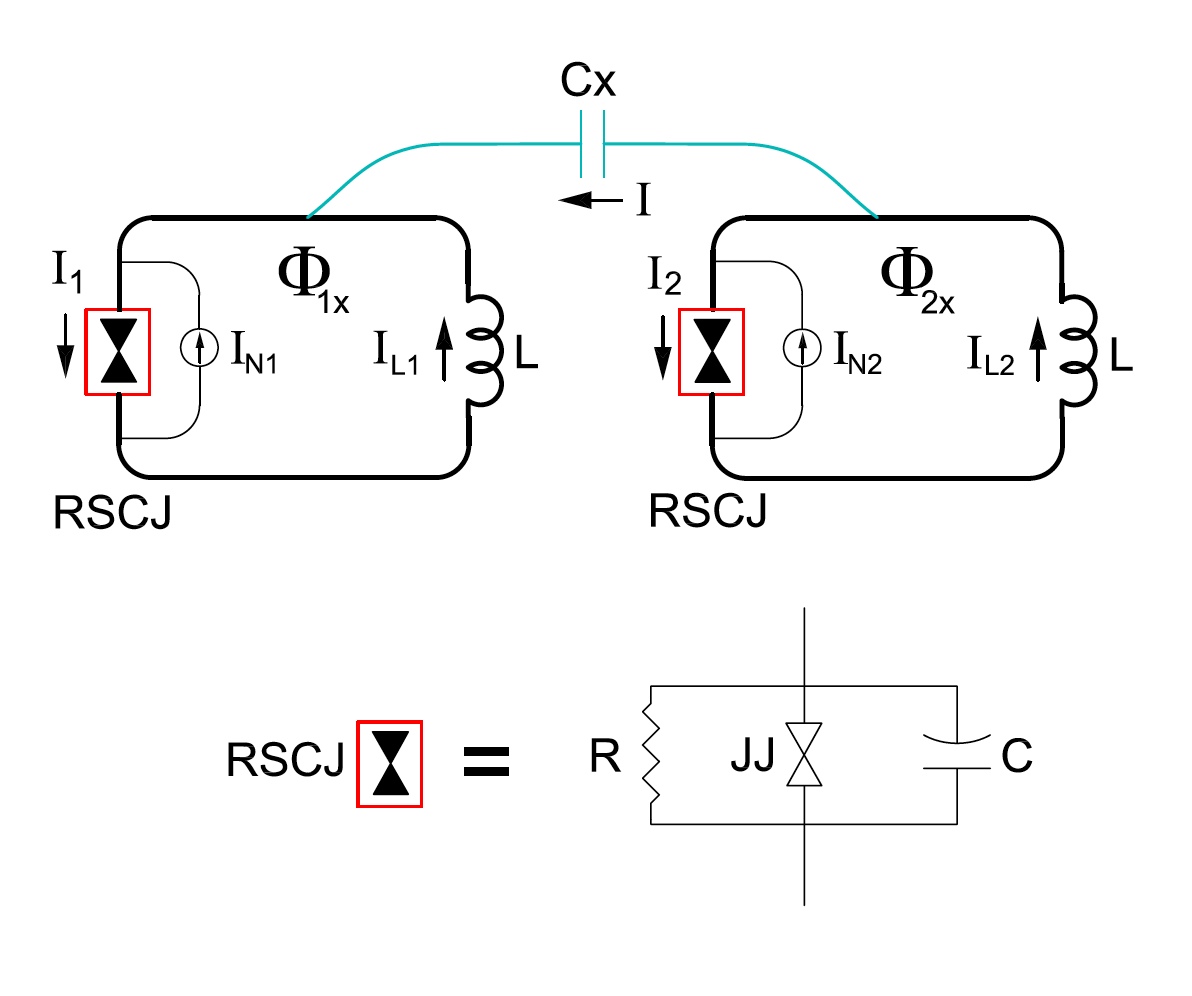}}
\end{center}
\vspace{-0.5 cm}
\caption{Equivalent circuit used in the
RCSJ simulations.}
\label{Fig17}
\end{figure}

For this treatment, we express this schematic as shown in Fig.\ref{Fig17}.

As
indicated, each junction is represented by its RCSJ equivalent: the parallel
combination of a critical current $I_{C}$, a resistance $R$ , and a
capacitance $C$. Each loop in the experimental sample has an associated
inductance $L$. $\Phi _{1x}$ is an external flux applied to loop \#$1$; $%
\Phi _{2x}$ is applied to loop \#$2$.

The phase dynamics for the two junctions depicted in Fig.\ref{Fig17} are
governed by the following equations.

\begin{eqnarray}
{\ddot{\varphi}_{1}+\alpha \dot{\varphi}_{1}+\sin \varphi _{1}} &{=\gamma
_{X}\left( \ddot{\varphi}_{2}-\ddot{\varphi}_{1}\right) -\beta
_{L}^{-1}\left( \varphi _{1}+2\pi M_{1x}\right) }&  \label{loops1} \\
{\ddot{\varphi}_{2}+\alpha \dot{\varphi}_{2}+\sin \varphi _{2}} &{=\gamma
_{X}\left( \ddot{\varphi}_{1}-\ddot{\varphi}_{2}\right) -\beta
_{L}^{-1}\left( \varphi _{2}+2\pi M_{2x}\right) }&  \label{loops2}
\end{eqnarray}%
where overdots denote derivatives in dimensionless time $\tau =\omega _{J}t$
and with each of the junction plasma frequencies being $\omega _{J}=\sqrt{%
2eI_{C}/\hbar C}$. The parameters were $\alpha =1/\omega _{J}RC$ and as
before $\beta _{L}=2\pi LI_{C}/\Phi _{0}$ with $\Phi _{0}$ being the flux
quantum. The two external fluxes appear in$M_{ix}=\Phi _{ix}/\Phi
_{0},\;\;\;i=1,2$ with the mutual coupling coefficient $\gamma _{X}=C_{x}/C$%
, $C_{x}$ is the macroscopic coupling capacitor, as shown in the figure.

\begin{figure}[t]
\begin{center}
\scalebox{0.35}{\centering \includegraphics{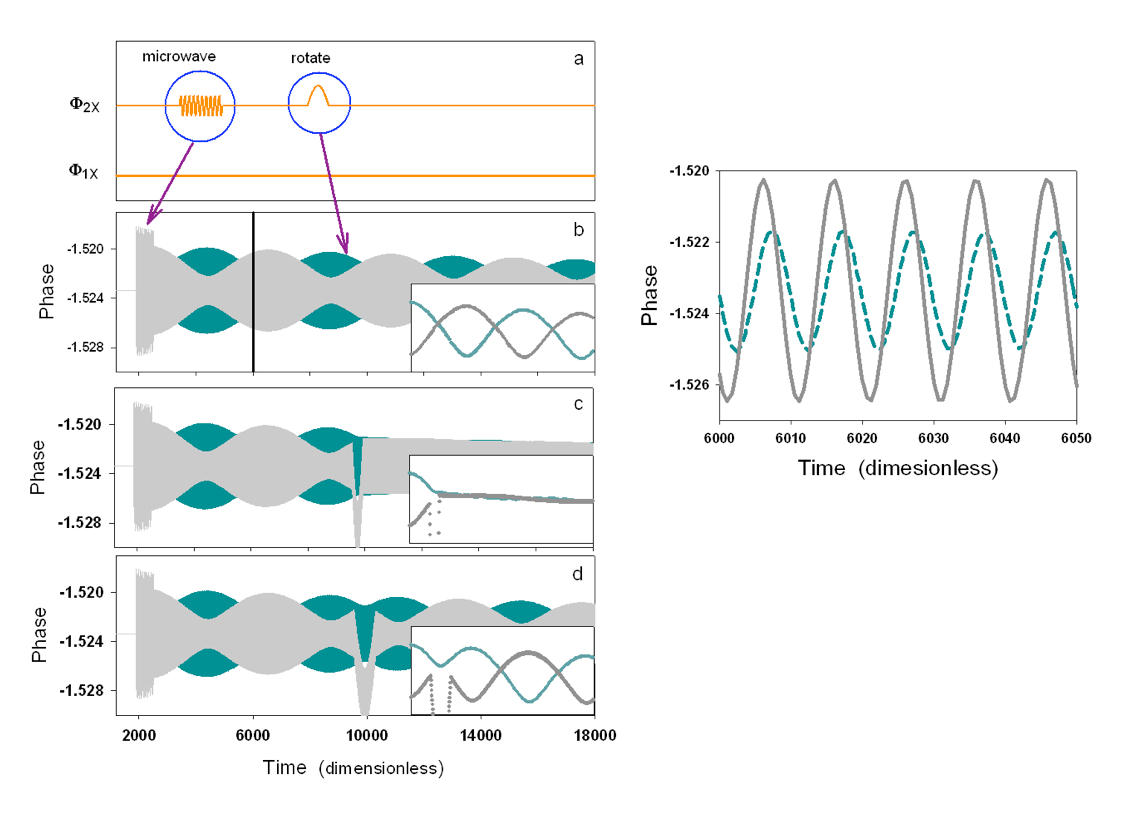}}
\end{center}
\vspace{-0.5 cm}
\caption{Left: RCSJ simulation
results for coupled loops. The external bias flux for each loop is depicted
in panel a. Panels b,c,d show the result of the applicationof a rotate pulse
of varying duration - 0, 60, and 120 plasma periods; Right: Phase
oscillationsexpanded around the vertical line in panel b of the figure on the
left.}
\label{Fig19}
\end{figure}

Based on published experimental data \cite{Steffen}, we set the parameters
at: $\alpha =5\times 10^{-5}$ (very light damping), $\beta
_{L}=2.841,\;\gamma _{X}=0.00231,\;I_{C}=1.1\mu A,\;C=1.3pF$, and $\omega
_{J}^{-1}=0.02ns$ (plasma period $12.6ns$). The critical applied flux was
thus $\Phi _{xc}=0.7022$; both loops were biased with a normalized dc flux $%
\Phi _{x}=0.6941$. Stimulus signals were added to the flux bias on one of
the loops as indicated in Fig.\ref{Fig19}, which also displays the results
of a numerical solution \cite{Blackburn4} of the coupled differential
equations. The duration of the microwave burst was set to $100$ plasma
periods; the amplitude and normalized frequency were $0.000580$ and $0.989$,
respectively. A rotation pulse of amplitude $0.000915$ was applied at time $%
9550$; its duration in each of the three panels was $0,\;60,$ and $120$
plasma periods, corresponding to $0,\;7.5,$ and $15$ ns.

This is a case of two coupled oscillators, one of which is kick-started with
an applied signal, after which there is a back and forth exchange of energy.
This is entirely analogous to the case of two weakly coupled pendulums -- a
standard topic in classical mechanics \cite{Chow}. In that situation the
pendulum oscillations exhibit low frequency envelopes. The undulations of
the junction phases depicted in the three lower panels in the figure on the
left are composed of rapid excursions of the individual phases between their
extremes (max and min); the close packing of the wave forms on that time
scale results in the appearance of a solid color fill. The insets in these
three panels show just the upper boundaries traced by the maxima of the
junction phases. The figure on the right has an expanded time scale (around
the vertical line in panel $b$) to reveal the uncompressed phase
oscillations. The ratio of the envelope period to the junction oscillation
period is more than $400:1$. In fact, as will be seen, it is this envelope
that is probed by experiments.

\begin{figure}[t]
\begin{center}
\scalebox{0.35}{\centering \includegraphics{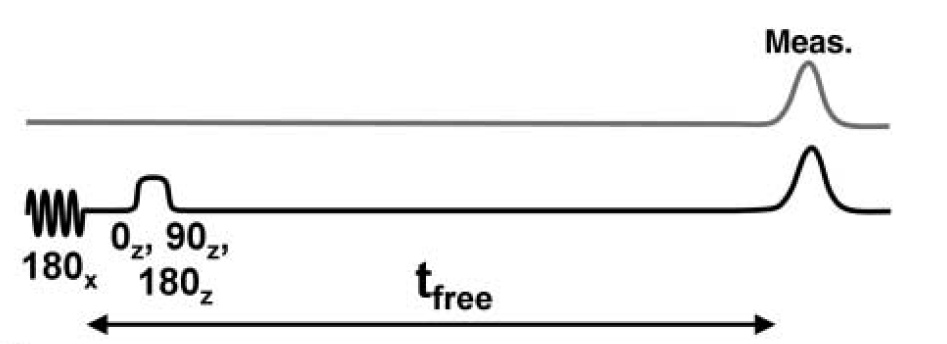}}
\end{center}
\vspace{-0.5 cm}
\caption{Bias pulse sequence as shown in
Fig. (1) in \protect\cite{Steffen}.}
\label{Fig20}
\end{figure}

To mimic the \textbf{experiment}, two additional elements must be added to
the previous treatment. First, the finite sample temperature ($25mK$) needs
to be included via the noise sources identified as $I_{N1}$ and $I_{N2}$ in
the circuit schematic. Second, measurement pulses are required as indicated
in this figure (\ref{Fig20}) taken from \cite{Steffen}.

\begin{figure}[t]
\begin{center}
\scalebox{0.3}{\centering \includegraphics{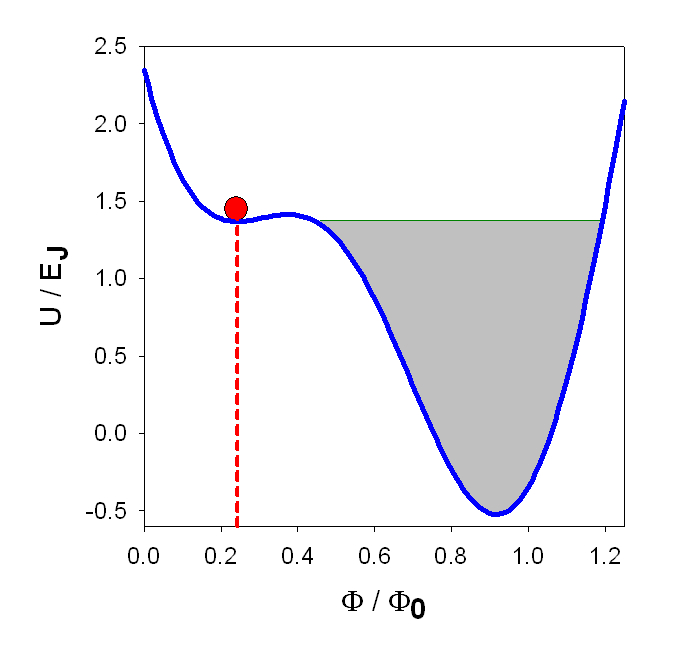}}
\end{center}
\vspace{-0.5 cm}
\caption{Energy-flux diagram for a bias
close to the critical external flux. \ If the "particle" escapes from this
shallow well, it will enter the much deeper well to the right and then
oscillate around the new minimum.}
\label{Fig21}
\end{figure}

To understand the action of a measurement pulse on either of the loops,
consider Fig \ref{Fig21} which shows a close up view of the potential energy
as a function of the net internal loop flux for an applied flux of $0.6941$
- just less than the critical value of $0.7022$. At this bias, the minimum
in the potential well is located at $\Phi/\Phi_{0}=0.242$.

The scenario goes as follows. Following the microwave pulse, the `particle'
in that loop is stimulated to oscillate at the bottom of its well. The
amplitudes of those oscillations, as depicted in the previous section, are
very small and centered around the minimum; they range from $0.2419<\Phi
/\Phi _{0}<0.2430$ and are not discernable at the scale of this plot.

In our simulation \cite{Blackburn4}, a triangular measure pulse of amplitude 
$0.00358$ was applied at a set free time after the oscillations commenced;
this is enough to briefly raise the applied flux to $0.6977$, closer to the
critical value, allowing noise to pop the `particle' over into the larger
parabolic well indicated by shading in Fig.\ref{Fig21}. Oscillations that
bring the `particle' closest to the barrier just before the measure pulse
stand the best chance of jumping out of the well with the aid of thermal
noise. Escape to the deeper well and subsequent back and forth oscillations
within it will be signalled by much larger phase oscillations and thus,
voltage oscillations.

An individual simulation run produces one of two outcomes following the
measure pulse: escape or non-escape. Repeated runs with the same time $%
t_{free} $ will yield a relative probability value for escape at that moment
and that in turn is a measure of the magnitude of the phase oscillations --
that is, the envelopes shown in the earlier figure.

The undriven but coupled loop will respond to the oscillations induced in
the driven loop.

To match the experiment, the simulation was carried out as follows. For any
chosen moment following the microwave burst, repeated simulation runs ($1000$
measurements) were carried out and data were gathered on how often there had
been an escape in each loop as indicated by a jump in voltage following the
measurement pulse.

\begin{figure}[t]
\begin{center}
\scalebox{0.35}{\centering \includegraphics{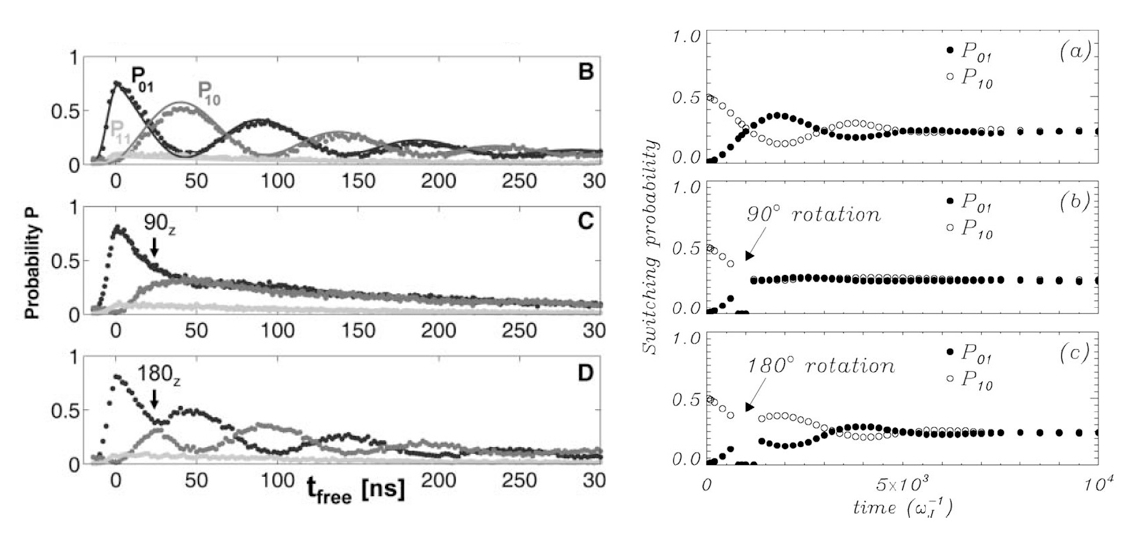}}
\end{center}
\vspace{-0.5 cm}
\caption{Left: Experimental
observations as shown in Fig.2 of \protect\cite{Steffen}. Each panel is for a different
rotation pulse; Right: RCSJ simulation results matching the corresponding
panels on the left.}
\label{Fig22}
\end{figure}

The results of the RCSJ-based simulation are shown in Fig.\ref{Fig22}
together with the experimental data from \cite{Steffen}.

\subsection{Entanglement of Two Qubits}

In \cite{Steffen} the devices were described as \textquotedblleft phase
qubits\textquotedblright . However, as was pointed out in the earlier
discussion, each Josephson junction was imbedded in a superconducting loop
of inductance $850pH$ and so the energy consisted of a part due to the
circulating supercurrent and a contribution from the Josephson coupling
energy. Therefore in this situation these are flux qubits, not phase qubits.
The loops are flux biased, not current biased. This is evident from Steffen 
\cite{Steffen} Fig.1A where the \textquotedblleft qubit
bias\textquotedblright\ is inductively coupled into the qubit loop. This
situation is distinctly different from the more common current-biased
Josephson junction where the energy takes the form of a \textquotedblleft
washboard potential\textquotedblright . In the quantum picture for this
system, it a fluxon that would escape from the well by tunneling out into
the larger and deeper well.

We remark that the RCSJ model suggests that the action of the microwaves is
merely to kick the `particle' into oscillations within its shallow well, so
in principle a brief plain pulse would work just as well. The MQT view is
that microwaves are necessary to pump the `particle' to a higher level in
that initial well. So a significant test of this issue would be to
substitute a simple pulse in both experiment and simulation. We would
predict the experimental data would not be affected.

\begin{figure}[t]
\begin{center}
\scalebox{0.35}{\centering \includegraphics{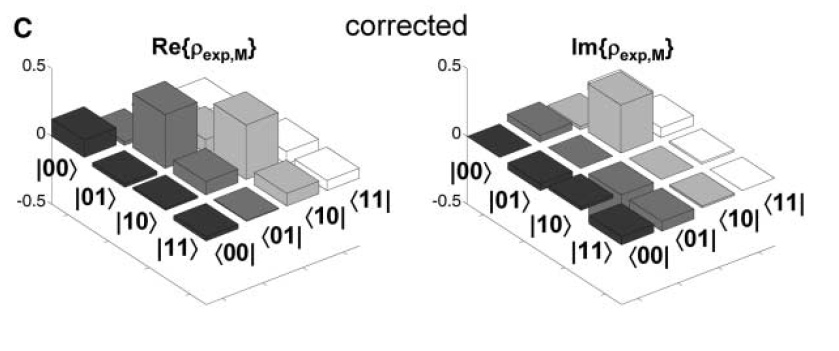}}
\end{center}
\vspace{-0.5 cm}
\caption{Figure 3c from \protect\cite%
{Steffen} showing state tomography applied to their experimental data.}
\label{Fig29}
\end{figure}

\begin{figure}[t]
\begin{center}
\scalebox{0.35}{\centering \includegraphics{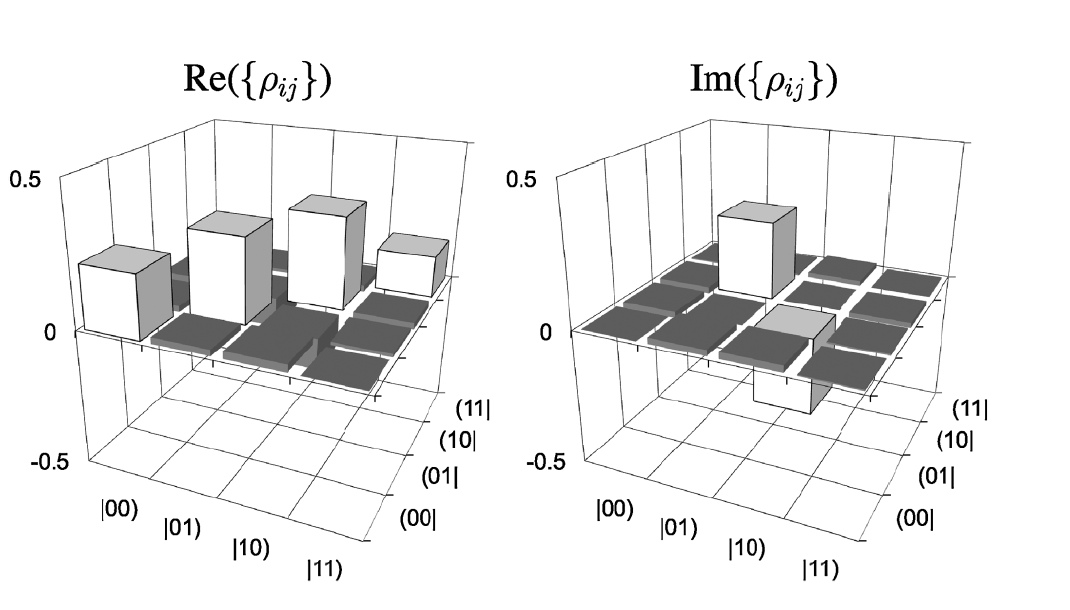}}
\end{center}
\vspace{-0.5 cm}
\caption{Results matching the previous
figure, but obtained from purely classical data generated by the RCSJ model.}
\label{Fig30}
\end{figure}

It is worth noting that in this case "unambiguous" evidence of quantum
behaviour of the coupled loops was claimed \cite{Steffen} based on the fact
that the density matrix had off-diagonal terms (see Fig.\ref{Fig29}) The
belief was that a classical system could not possess such a property. That
conjecture is naturally wrong if we think in terms of classical nonlinear
systems,\textbf{\ }and indeed simulations of rf-SQUIDs coupled through a
capacitor showed that a suitably defined density matrix could have
off-diagonal terms \cite{NGJ5}. These simulations returned a picture very
close to that obtained in the experiments, as shown in Fig.\ref{Fig30}. Bear
in mind that Fig. \ref{Fig29} derives from experimental data while Fig. \ref%
{Fig30} is obtained from numerical simulation data from a fully classical
model.\ Therefore state tomography cannot be considered a definitive test of
quantumness in the source of the data.

\textbf{\ }

\section{\protect\large Historical Remarks}

At first the idea of a classical/quantum transition became the subject of
experiments by just a few groups, and the evidence did seem to be consistent
with the macroscopic quantum hypothesis. Then the level of activity on the
topic rose significantly. Josephson systems such as rf-SQUIDs, large area
junctions, annular junctions, and configurations of coupled junctions were
all examined with the expectation that they would provide confirmation at
the lowest temperatures of the new quantum phenomenology, and most of the
obtained results seemed to support that idea. By the early 90's, in the
absence of any detailed classical RCSJ simulations, it was was generally
accepted within the community that the quantum hypothesis had been confirmed
by experiments. Like supersolids \cite{Hallock}, once the idea of a
macroscopic quantum state was out there, people pursued it.

Unfortunately, several aspects concerning RCSJ simulations were not clear in
the 80's, mainly because computer simulations for the nonlinear circuit
model, including dc and ac forcing currents, along with noise current terms,
represented a rather challenging topic and few groups around the world
having access to powerful computing facilities were able to tame it. Also,
at that time, the exponential growth of the interest for chaotic dynamics
moved the attention of the international community toward the intriguing and
interesting evidences of chaos in Josephson systems \cite{Kautz1} and, more
in general, toward spatio-temporal structures competition in nonlinear
systems \cite{Christ1}. Thus, at the beginning of the 90ies, in the absence
of specific and systematic RCSJ simulations, the existence of a crossover
temperature, of energy levels in the Josephson potentials and their evidence
through \textquotedblleft spectroscopy\textquotedblright\ were generally
accepted as a necessity.

The above historical conditions gave birth, at the beginning of the 90ies,
to a research project \cite{MQCNY}, aiming at securing evidence of
Macroscopic Quantum Coherence (MQC was the acronym of the project) along the
philosophy of a gedanken experiment proposed by Claudia Tesche \cite%
{TESCHE90}. The project was aiming to take advantage of the possible
macroscopic quantum behaviour of Josephson circuits in order to provide
proof of fundamental relations of quantum mechanics such as Bell's
inequalities. The experiment was based in particular on the double-well
potential of an rf-SQUID, namely a superconducting loop interrupted by a
Josephson junction. One well of the potential was supposed to be occupied by
a current circulating clockwise while the other well was occupied by a
current circulating counterclockwise. In principle, if the barrier between
the two wells was sufficiently low, the net current in the rf-SQUID loop
could be considered as being a superposition of the state of each well, and
the system itself could be in a typical quantum singlet state. Using a
dc-SQUID coupled to an rf-SQUID that served as a detector for the
superposition state ,it should have been possible to probe the quantum
properties of the superposition itself (see Fig.2 \cite{MQCNY}). The MQC
experiment resulted in the development of a thermal escape technique
necessary for probing tunnelling and potentials\cite{Kautz,NGJ0,Barb}.

At some point the basic circuit architecture of the MQC experiment was
changed in order to introduce, following an idea of J. E. Lukens and
co-workers \cite{LUPRB92, LUNA00}, a means of tuning of the height of the
barrier separating the two wells of the rf-SQUID potential. A lower barrier
could favor the interaction between the macroscopic quantum states in the
two wells while the tunability could pinpoint the moment when the quantum
interaction turned on. This tunability could be realized by replacing the
single junction of the rf-SQUID loop with a two parallel junctions in a
small subsidiary loop as shown in Fig.\ref{tunablebarrier}. 
\begin{figure}[t]
\begin{center}
\scalebox{0.25}{\centering \includegraphics{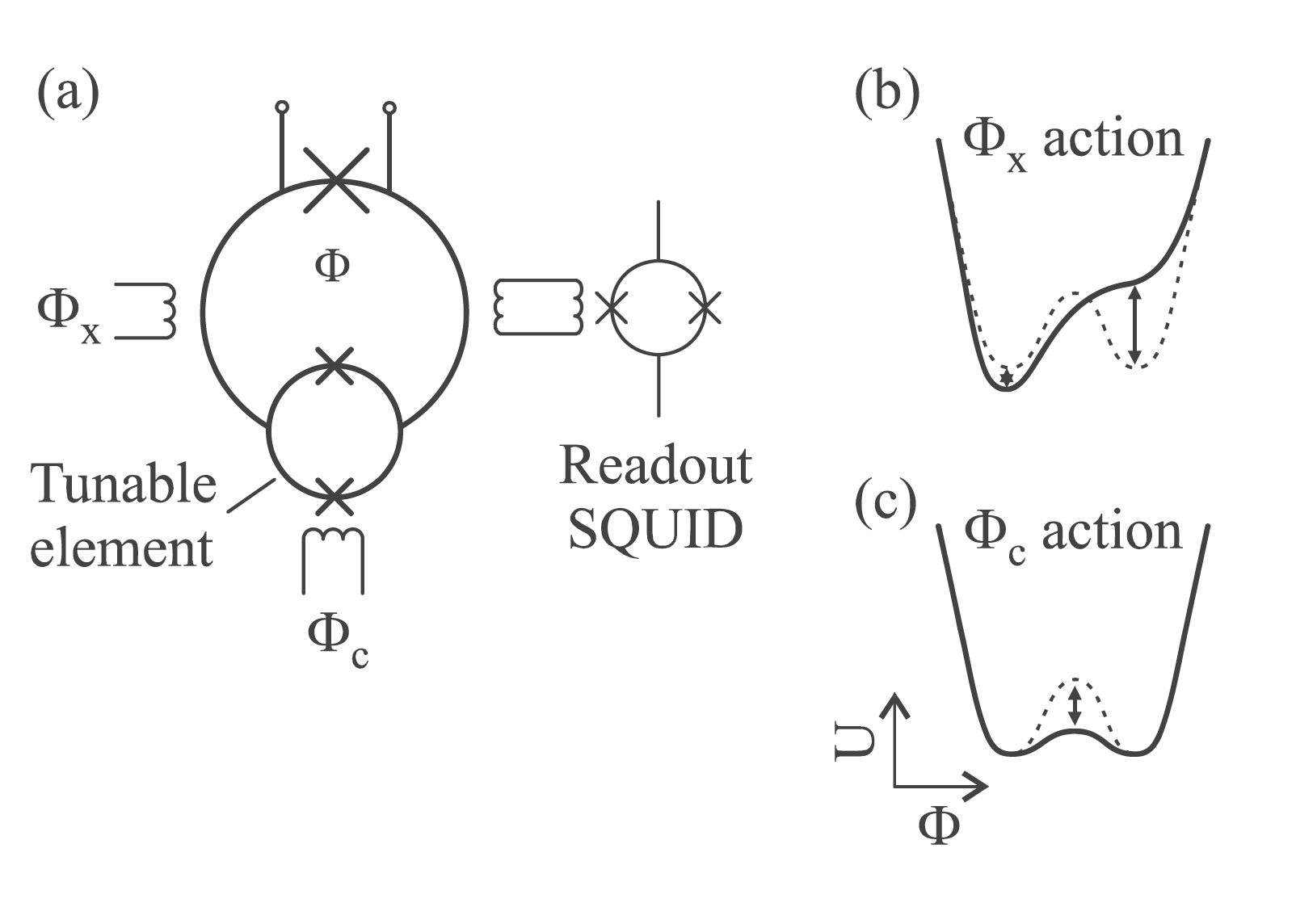}}
\end{center}
\vspace{-0.5 cm}
\caption{An rf squid containing a tunable element - two
parallel junctions in a small sub- loop (dc squid).}
\label{ftunablebarrier}
\end{figure}
Sophisticated and sensitive
potential escape measuremets \cite{CATA07} at very low temperatures showed
that the role of the inner loop giving rise to the tunability of the system
modifies the shape of the potential giving rise to interesting features
which can be explained within the general frame of the analysis of
singularities in classical nonlinear systems \cite{Thom75}. It is most
likely the nonlinear dynamic interaction between the junctions of the dc
interferometer which originates a \textquotedblleft gap\textquotedblright\
in the excitation spectrum which was interpreted as evidence of a quantum
effect due to level quantization \cite{LUNA00}: in general indeed, we have
found that a gap structure can arise in the excitation spectrum of two RCSJ
oscillators whenever a coupling between those is allowed.

Over all, interactions between the two wells of the rf-SQUID potential
expected in zero-applied microwaver (rf currents) have not been observed
from escape measurements. Some interesting observations have been recently
reported by pulsing the system with microwave burst \cite{Spilla14} ,
however, the discussion reported in the previous sections concerning the
"atom-like" responses of Josephson junctions indicates that collecting
information by pumping withy external microwaves, or short microwave bursts,
or anything else can hardly lead to unique \textquotedblleft
quantum\textquotedblright\ properties for Josephson systems . It must always
be recalled that the junctions used for macroscopic quantum coherence
experiments are of very good quality and, speaking in terms of RCSJ model,
very lightly damped, and therefore a short burst of microwaves, or even just
a single pulse, induces oscillations having a very long decay time which can
allow modelling and manipulations even from a classical point of view.

\section{\protect\large Recent Development}

\subsection{Qubits Coupled via Coax Cable}

In the first publication reporting experimental results for coupled qubits 
\cite{Steffen}, the coupling was provided by a macroscopic external
capacitor. The interpretation placed on the experimental findings was
entirely within a quantum narrative and concluded that entanglement had been
observed. A classical equivalent circuit based on the RCSJ model led to a
pair of coupled differential equations for the phase variables; numerical
solutions of these equations \cite{Blackburn4}, \cite{NGJ5} were shown to
fully replicate the experimental observations. \ This cast significant doubt
on the claims that quantum entanglement had been unambiguously demonstrated.
\ 

\begin{figure}[t]
\begin{center}
\scalebox{0.25}{\centering \includegraphics{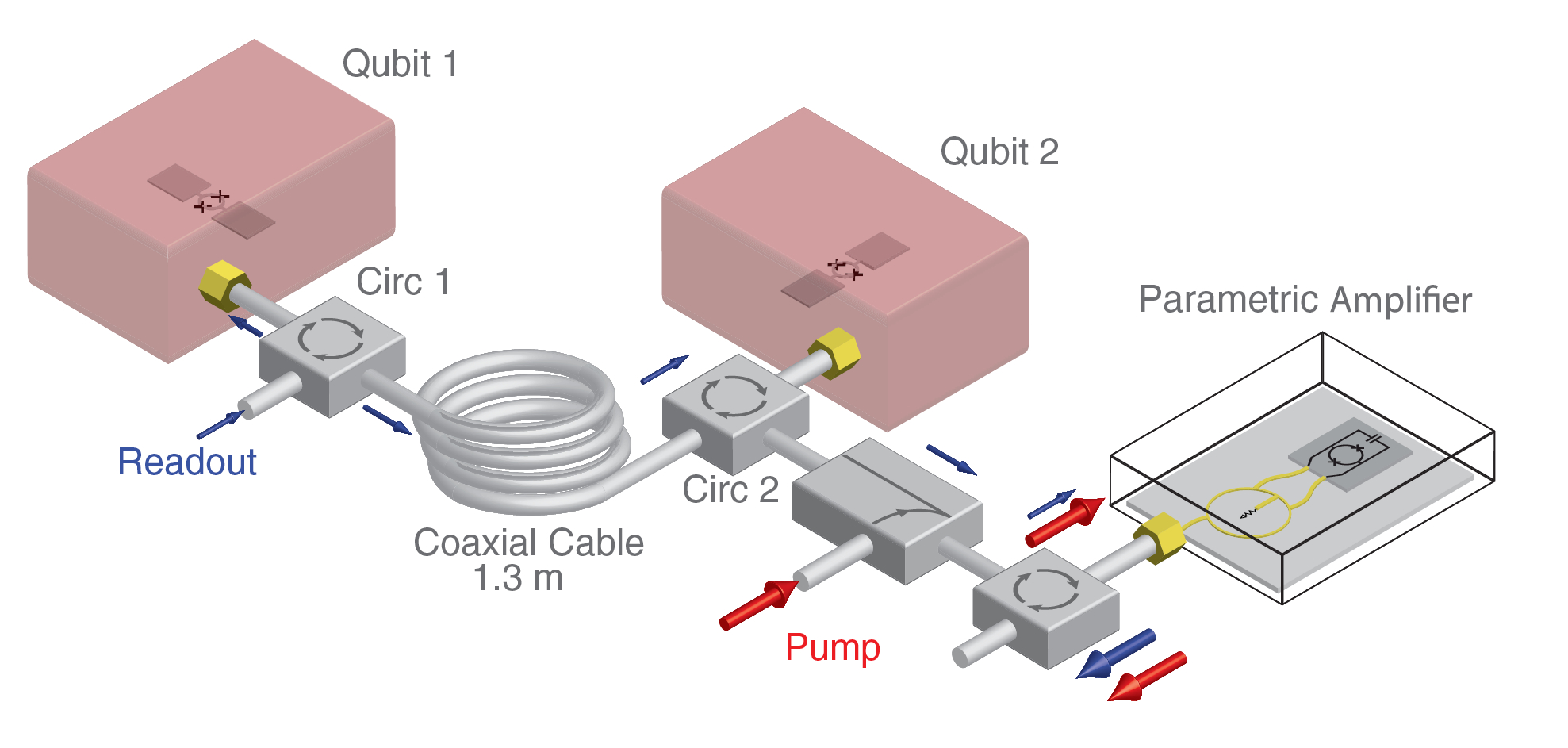}}
\end{center}
\vspace{-0.5 cm}
\caption{Depiction of coaxial cable coupling a pair of Josephson qubits.}
\label{coax}
\end{figure}

More recently, this kind of macroscopic coupling, now of two transmons, has
been extended from a capacitor to a one-meter transmission line, as shown in
Fig.\ref{coax} from \cite{Roch}.

\begin{figure}[t]
\begin{center}
\scalebox{0.25}{\centering \includegraphics{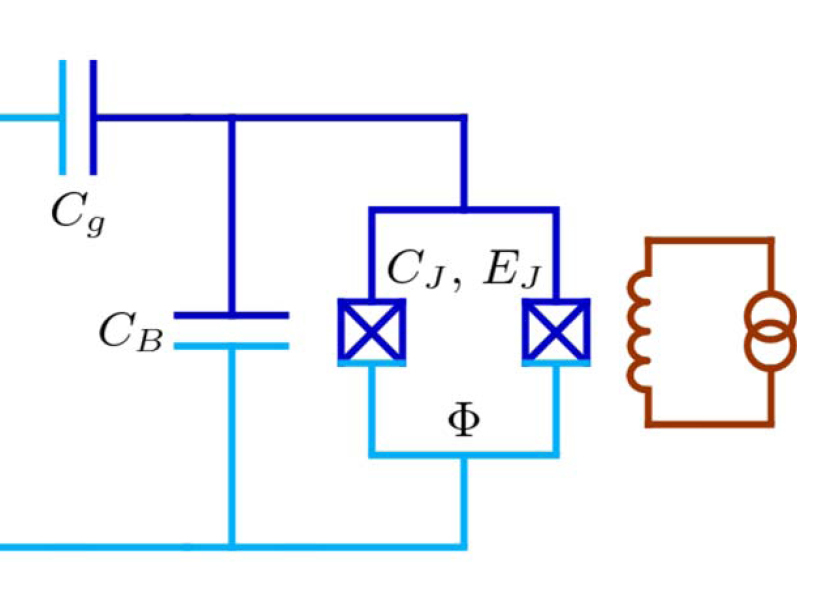}}
\end{center}
\vspace{-0.5 cm}
\caption{Equivalent circuit of a transmon
qubit. \ Each crossed-box is an RCSJ Josephson junction. The loop is flux
biased by the subcircuit to the right.}
\label{Koch Transmon}
\end{figure}

A transmon is a particular configuration of Josephson junctions, as depicted
in Fig. \ref{Koch Transmon}, taken from \cite{Koch} (see also \cite{Krantz}).

\section{\protect\large Final Comments}

All of the experiments reviewed here were intended either to confirm the
existence of a macroscopic quantum state for Josephson junctions at very low
temperatures, or to exploit the anticipated properties of a junction in the
macroscopic quantum state. The simplest and most straightforword
investigations were structured as swept-bias experiments, both with and
without applied microwaves. These experiments recorded the value of the bias
current at the moment when a finite voltage suddenly appeared across the
junction. \ Accumulated data from repeated bias scans provided switching
current distributions (SCD).\ Without microwaves, a single SCD peak is seen.
MQT theory predicted that the position of this peak should become
temperature independent at sufficiently low temperatures. Such peak freezing
is a \textit{sine qua non} of any claims that a macroscopic quantum state
has been achieved. As shown in this review, close inspection of experimental
data clearly indicates that peak freezing has not been seen.

When microwaves are present, additional SCD peaks appear, commonly just one
or two. These are well described as resulting from classical resonant
activation at the anharmonic, or harmonic, plasma frequency. What has not
been reported are multiple peaks expected from different pairs of
inter-level spacings in a well.

Next were experiments involving more complex superconducting circuits; they
were focused on expectations arising from the artificial atom picture of a
low temperature Josephson junction. Finally, there were experiments
consisting of a coupled pair of superconducting loops, each containing a
Josephson junction. \ These were intended to probe anticipated quantum
entanglement.

It is worth recalling that above a hypothetical transition temperature,
circuits containing Josephson junctions can be fully modelled with classical
components (capacitors, inductors, resistors) together with the
Johnson-McCumber-Stewart Josephson junction model from 1968. \ The key
question is whether a transition to a macroscopic quantum state occurs at
sufficiently low temperatures. At present the allure of a superconducting
quantum computer is undeniable, however, as we have seen in this Report,
there may have been a rush to judgment because, on close examination, much
of the collected experimental data did not, in a \textit{definitive}
fashion, conclusively support the hypothesis of a transition to macroscopic
quantum state.\ This applies to phase qubits, flux qubits, SQUID's, etc. \
Even some recent qubit configurations such as transmons may be expressed in
terms of the RCSJ model \cite{Koch} (see also \cite{Krantz}).

Beginning with the early experiments of Voss and Webb \cite{VossWebb}, and
Devoret et al.\cite{Devoret}, the accumulating evidence seemed to support
the conjecture that Josephson junctions at sufficiently low temperatures
would enter a new macroscopic quantum state. If true, Josephson junctions
could function as of superconducting quantum bits (qubits). The evidences
documented in this Review show that the interpretation of such experiments
in terms of the new macroscopic quantum states is far from unique and/or
necessary. A decisive experiment would be required whose results could 
\textit{only} be understood in quantum terms and to make this crucial step,
the RCSJ model must be applied and shown to fail.\bigskip

{\Large Acknowledgements}\textbf{:} \ This work was supported by GRF grant
\#235550 from Wilfrid Laurier University. We thank Rudolf Gross for
permission to use Fig.4 and Roberto Ramos for permission to use Fig.22 (left
panel). We also thank Massimiliano Lucci and Ivano Ottaviani for providing
Fig. 1 and Fig. 2.

Several colleagues suggested that we should undertake this endeavour in
order to combine all the relevant material into a single compact document.
We thank these colleagues for their encouragement, hoping that the resulting
effort will satisfy/motivate interest and curiosity in this fundamental
topic.

\section{Appendix}

\begin{figure}[t]
\begin{center}
\scalebox{0.5}{\centering \includegraphics{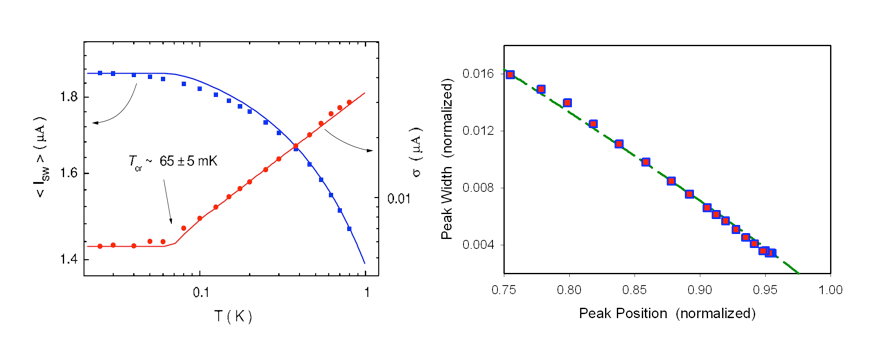}}
\end{center}
\vspace{-0.5 cm}
\caption{Left panel: Fig.2 from \protect\cite{Yu} showing
experimental SCD peak posions (blue squares) and peak widths (red circles)
as functions of temperature; Right panel: Dashed line - SCD peak positions
versus peak widths from simulations described in II-C, and the
experimental data shown in the left panel  (blue/red squares).}
\label{Fig31}
\end{figure}

At the very beginning of the MQT story, in 1981, the paper by Voss and Webb 
\cite{VossWebb} gave pride of place to SCD peak \textit{width} over peak 
\textit{position}. In addition, peak width data were presented with
logarithmic temperature scales which, as has been shown in this Report,
tends to visually suggest a saturation of escape behavior at the lowest
temperatures. This format for showing experimental escape data became a
defacto standard. But in actual fact, position and width are interdependent;
as the peak position goes up, the peak width decreases. Figure \ref{Fig31} \
shows this relationship between width and position produced from the RCSJ
simulation discussed in Part I. Also shown are experimental data points
taken from Yu et al. \cite{Yu}, Fig 2. This confirms, both theoretically and
experimentally, that the peak width and peak position are directly related.
Therefore neither parameter contains more information than the other. Yet,
the peak position would have been a better choice for presenting data since
in the lowest temperature region the peaks become exceedingly narrow,
meaning the widths become more and more difficult to determine
experimentally.

Therefore the most appropriate consideration of experimental data would use
peak positions and a linear temperature scale

\end{document}